\documentstyle[12pt,aaspp4,flushrt]{article}

\received{}
\accepted{}

\slugcomment{}

\lefthead{Bower et al.}
\righthead{Nucleus of NGC 1023}

\begin{document}

\title{Evidence of a Supermassive Black Hole in the Galaxy NGC 1023
from the Nuclear Stellar Dynamics\altaffilmark{1}}

\author{G.~A.~Bower\altaffilmark{2}, R.~F.~Green\altaffilmark{2}, 
R.~Bender\altaffilmark{3}, K.~Gebhardt\altaffilmark{4}, T.~R.~Lauer\altaffilmark{2},
J.~Magorrian\altaffilmark{5}, 
D.~O.~Richstone\altaffilmark{6},
A.~Danks\altaffilmark{7},
T.~Gull\altaffilmark{8}, J.~Hutchings\altaffilmark{9},
C.~Joseph\altaffilmark{10}, M.~E.~Kaiser\altaffilmark{11}, 
D.~Weistrop\altaffilmark{12}, B.~Woodgate\altaffilmark{8},  
C.~Nelson\altaffilmark{12}, and E.~M.~Malumuth\altaffilmark{7}}

\altaffiltext{1}{Based on observations with the NASA/ESA {\it Hubble Space Telescope},
obtained
at the Space Telescope Science Institute, which is operated by the
Association of Universities for Research in Astronomy, Inc.~(AURA), under
NASA contract NAS5-26555.}

\altaffiltext{2}{Kitt Peak National Observatory, National Optical Astronomy Observatories, P.~O.~Box 26732,
Tucson, AZ 85726; gbower@noao.edu, operated by AURA, Inc.~under cooperative agreement with 
the National Science
Foundation.}
\altaffiltext{3}{Universit\"{a}ts-Sterwarte, Scheinerstrasse 1, D-81679 M\"{u}nchen, Germany.}
\altaffiltext{4}{Department of Astronomy, University of Texas at Austin, Austin, TX 78712.}
\altaffiltext{5}{Institute of Astronomy, Cambridge University, Madingley Road, Cambridge,
CB3 0HA, England.}
\altaffiltext{6}{Department of Astronomy, University of Michigan, Dennison Building, Ann
Arbor, MI 48109.}
\altaffiltext{7}{Raytheon ITSS, NASA/Goddard Space Flight Center, Code 681, Greenbelt, MD 20771.}
\altaffiltext{8}{NASA/Goddard Space Flight Center, Code 681, Greenbelt, MD 20771.}
\altaffiltext{9}{Dominion Astrophysical Observatory, 
National Research Council of Canada,
5071 W.~Saanich Road, 
Victoria, BC V8X 4M6, Canada.}
\altaffiltext{10}{Dept.~of Physics \& Astronomy, Rutgers University,
P.~O.~Box 849, Piscataway, NJ 08855.}
\altaffiltext{11}{Department of Physics \& Astronomy, Johns Hopkins University,
Homewood Campus, Baltimore, MD 21218.} 
\altaffiltext{12}{Department of Physics, University of Nevada, 4505 S.~Maryland
Parkway, Las Vegas, NV 89154.}

\newpage

\begin{abstract}

We analyze the nuclear stellar
dynamics of the SB0 galaxy NGC~1023, utilizing observational data both
from the Space
Telescope Imaging Spectrograph aboard the Hubble Space Telescope and from the ground.
The stellar kinematics measured from these long-slit spectra show rapid rotation
($V \approx 70$ km s$^{-1}$ at a distance of $0\farcs1 = 4.9$ pc from
the nucleus) and increasing velocity dispersion toward the nucleus (where 
$\sigma = 295 \pm 30$ km s$^{-1}$).  
We model the observed stellar kinematics assuming an axisymmetric mass distribution with
both two and three integrals of motion. Both modeling techniques point to the
presence of a central dark compact mass (which presumably is a supermassive black hole)
with confidence $> 99$\%.
The isotropic two-integral models yield a best-fitting
black hole mass of $(6.0 \pm 1.4) \times 10^7 \ M_{\odot}$ and mass-to-light ratio 
($M/L_V$)
of $5.38 \pm 0.08$, and the goodness-of-fit ($\chi^2$) is insensitive to 
reasonable values 
for the galaxy's
inclination. The three-integral models, which non-parametrically
fit the observed line-of-sight velocity
distribution as a function of position in the galaxy, suggest a black
hole mass of $(3.9 \pm 0.4) \times 10^7 \ M_{\odot}$ and $M/L_V$ of $5.56 \pm 0.02$
(internal errors), 
and the
edge-on models are vastly superior fits over models at other inclinations.
The internal dynamics in NGC~1023 as suggested by our best-fit three-integral
model shows that the velocity distribution function at the nucleus is
tangentially anisotropic, suggesting the presence of a nuclear stellar disk.
The nuclear line of sight velocity distribution has enhanced wings at
velocities $\geq 600$ km~s$^{-1}$ from systemic, suggesting that perhaps 
we have detected a group
of stars very close to the central dark mass.

\end{abstract}


\newpage   

\section{Introduction}

Supermassive black holes (BHs) are widely suspected to be the central engines of
quasars and active galactic nuclei. The fact that the comoving number density of
quasars peaks at $z \sim 2 - 3$ (e.g., Schmidt, Schneider, \& Gunn~1995) and declines by
nearly three orders of magnitude by $z = 0$ implies that nearby galaxies should harbor
relatively inactive BHs. Searching for these
BHs would provide 
constraints on the evolutionary history of quasars.
Also, the most recent census of nearby BHs reveals a remarkable correlation
between BH mass and the stellar velocity dispersion measured within the half-light
radius of its host galaxy (Gebhardt et al.~2000a;
Ferrarese \& Merritt 2000). This result will have very interesting 
implications
for the formation of galaxy spheroids and the growth of their central BHs (Kormendy 2000).
 
Dynamical evidence of dark
compact objects of mass $\sim 10^{6.5} \ - \ 10^{9.5} \ M_{\odot}$ has been accumulating
rapidly in the last decade or so (see reviews in Kormendy \& Richstone~1995; Richstone
et al.~1998). Although this evidence in most cases does not strictly exclude 
alternatives to
a BH, it is very difficult to construct plausible alternative models to BHs for three of
these BH candidates (the Galaxy, NGC 4258, and M31; see Maoz~1998 and Kormendy \& 
Bender~1999) $-$ thus suggesting that many, or perhaps all, of the dark compact objects
that have been identified through such dynamical evidence are indeed BHs.

The Hubble Space Telescope (HST)
has greatly enhanced this effort to analyze 
the nuclear dynamics of nearby galaxies.  
HST's second generation spectrograph,
the Space Telescope Imaging Spectrograph (STIS; see Woodgate et al.~1998) 
can obtain measurements
of the nuclear kinematics of either gas or stars with
much more efficiency than the first generation HST spectrographs.

Several teams are actively involved in this research.
Our group, a subset of the STIS Investigation Definition Team, has selected a
galaxy sample including both some of the best black hole candidates and a few
galaxies where the ground-based dynamical evidence for a BH is
nonexistent or weak.
One such galaxy in the latter category is NGC~1023, 
which is classified as SB(rs)0$-$. Given this morphological type and that $M_B = -19.9$, 
the luminosity of the bulge 
component is $M_{B,bulge} \approx -19.4$ 
(see Simien \& deVaucouleurs~1986).
We chose NGC 1023 because its intermediate luminosity is consistent with a rotationally
supported figure (Faber et al.~1997), removing ambiguity from the interpretation of the 
nuclear dynamics.

We adopt a distance
of 10.2 Mpc (see Faber et al.~1997); thus $1''$ corresponds to a distance at NGC~1023 of
49 pc. The Galactic extinction is $A_B = 0\fm262$ (Schlegel et al.~1998).

\section{Observations and Data Calibration}

Analyzing the nuclear stellar dynamics of galaxies requires imaging and
spectroscopy using both HST and ground-based observations. HST observations
measure the galactic central structure and stellar kinematics within a distance 
of a few arcseconds of
the nucleus where the radial gradients in intensity and kinematics are steepest.
Ground-based observations extend the spatial coverage to large radius so that the
global photometric and kinematic structure of the galaxy can be fitted during
construction of galaxy dynamical models.

\subsection{Surface Photometry}

Continuum images of the nuclear region of NGC~1023 were obtained 
with 
the Wide Field Planetary
Camera 2 (WFPC2) aboard HST on 1996 January 27 by Lauer et 
al.~(2000). Since they will present the details of the imaging analysis, here we will
summarize only the details relevant to NGC~1023.
The nucleus was placed in the Planetary Camera, which has a plate scale of $0\farcs0455$
pixel$^{-1}$. Several exposures were obtained through each of the filters F555W and F814W,
which are roughly equivalent to the V and I filters. The total signal attained at the nucleus
was roughly $22000$ $e^-$ pixel$^{-1}$ and $36000$ $e^-$ pixel$^{-1}$, respectively. 
The images were reduced and deconvolved with 40 iterations of the Lucy-Richardson
method (Lucy 1974; Richardson 1972).
We measured the surface photometry in Fig.~1 
using isophote fits as described in Lauer et al.~(1995). We need to find a fit to this
surface brightness profile because in \S 3.2 we will examine the effects of the STIS
PSF. First, we fit
a Nuker law (Lauer et al.~1995; Faber et al.~1997):

\begin{equation}
I(r) = 2^{(\beta - \gamma)/\alpha} I_b \left ({r_b \over r} \right )^\gamma
\left [ 1 + \left ({r \over r_b} \right )^\alpha \right ]^{(\gamma - \beta)/\alpha},
\end{equation}

\noindent
where $I_b$ corresponds to a surface brightness of 16.05 mag/$''^2$, $r_b =
1\farcs64$, $\alpha = 2.14$, $\beta = 1.26$, and $\gamma = 0.70$. This profile is
a steep, largely featureless power law, characteristic of the profiles in early-type
galaxies with modest bulge luminosity (Faber et al.~1997). The central two points
in the profile were excluded from the Nuker law fit, since this law does not apply
to such sharp increases in intensity right at the nucleus. Thus, we confirm the 
detection of the
compact nuclear component (presumably a nuclear stellar cluster) in NGC~1023 by
Lauer et al.~(1995). To fit the nuclear star cluster, we added the following function
to the fitted Nuker law:

\begin{equation}
I(r) = {I_0 \over \left [ 1 + (r/r_0)^2 \right ]^n},
\end{equation}

\noindent
where $I_0 = 11.69$ mag/$''^2$, $r_0 = 0\farcs035$, and $n = 2.0$.
This fit was constrained with the inner three data points in the
profile and the total flux in the STIS spectrum (\S 2.2) within $0\farcs075$ of the nucleus
[given an adopted (V$-$I) of 1.4]. The profile of the nuclear cluster cannot be much
steeper without violating at least one of these constraints.
We use this fit to
the observed profile only in \S 3.2.
In our analysis in \S 4 where we construct dynamical models, 
we use the photometric data instead. 

To extend the photometry to larger radius, we added the photographic photometry of
NGC~1023 from Barbon \& Capaccioli (1975). Their photographic plates show a `cloud'
$\sim 2\farcm7$ east of the nucleus, which is presumably an individual dwarf
galaxy perhaps in the process of merging with NGC~1023.
Since our kinematic data discussed
below extend out to only $\sim 100''$ from the nucleus, our galaxy dynamical 
models (\S 4) are only
applicable inside this radius. Therefore, we ignore this dwarf galaxy by adopting
Barbon \& Capaccioli's surface brightness data for the west side of the galaxy
(away from the companion dwarf galaxy).
We verified Barbon \& Capaccioli's (1975) surface photometry by comparing it to
the intensity along the slit from the ground-based spectroscopy described
in \S 2.2 near
5200 \AA \ from Gebhardt et al.~(2000b). 
In order to combine B-band photographic
surface photometry with the V-band HST surface photometry, we normalized their
photographic data by adopting (B$-$V) = 1.0 (de Vaucouleurs et al.~1991).

\subsection{Long-slit Spectroscopy}

The primary set of ground-based long-slit spectroscopic observations of NGC~1023
will be presented in Gebhardt et al.~(2000b). 
For these observations, their $1''$ wide slit was aligned along the three position 
angles
(P.A.) of $0\arcdeg$, $90\arcdeg$, and $45\arcdeg$. The first two angles correspond to
the photometric minor 
and major
axes. The spatial resolution (dominated by atmospheric seeing) was
$\approx 1\farcs0$ (FWHM).
The spectra covered 4765 $-$ 5788~\AA, which includes the absorption lines 
of Mg I b (5167~\AA, 5172~\AA, and
5183~\AA), with a resolution (FWHM) of 3.1~\AA.  

Since the measurement of stellar kinematics is usually easier using the Ca II triplet 
(8498~\AA, 8542~\AA, and 
8662~\AA) 
absorption lines compared to Mg I b, we 
obtained additional long-slit spectra of NGC~1023 
with the KPNO 4 m telescope covering the Ca II triplet. Table~1 lists the
details of our KPNO observation, including the detector employed, wavelength
coverage, slit size, and spectral resolution.
We observed at the
P.A.'s of $0\arcdeg$ and $90\arcdeg$. 
The atmospheric seeing was
$1\farcs8 - 2\farcs2$ (FWHM).
The nucleus was placed in the
slit, as viewed through the telescope guide
camera, prior to the start of each galaxy exposure.
We eliminated the
possibility of centering errors caused by differential atmospheric
refraction between the guide camera and the spectrograph slit, by
mounting an
I-band filter in front of the guide camera.
Galaxy exposures were typically 1800 sec in duration, after which
a spectrum of the comparison lamp was
obtained. The spectrum of a nearby faint star was then
obtained to measure the seeing-dominated PSF in the spatial
direction.
The galaxy nucleus
was then carefully moved along the slit by
$\approx 5$ pixels, and another exposure was
taken. We repeated this exposure sequence for $2-4$ hours of integration time per slit
P.A.; the $S/N$ in the reduced spectra is $\approx 100/$\AA \ at the nucleus.
Spectra of bright K giant stars (to serve as template spectra during the analysis)
were obtained during twilight. 
The spectra were calibrated using standard techniques in IRAF for
optical long-slit spectroscopy. Wavelength calibration was established from the positions of
several emission lines in the comparison spectra, providing
for wavelength calibrations accurate to $\pm 0.08$~\AA. The continuum peaks
in the spatial profile of each galaxy exposure were aligned before combining the 
individual spectra.

Now we describe our observing procedure using HST/STIS.
To obtain a stellar template spectrum needed for measuring the stellar kinematics in a
galaxy STIS spectrum, the K0~III star HR~7615 was observed with the STIS CCD on 1997 
October 
23. The instrumental setup was identical to that of
the NGC~1023
visit described below. Spectra were obtained with the star at each of five positions
along the slit width (i.e., along the dispersion axis), corresponding to $-0\farcs10$,
$-0\farcs05$, $0\farcs00$, $+0\farcs05$, and $+0\farcs10$ (where the origin is at the
slit center). The star was stepped along the slit in this way so that during data analysis
we could combine the individual template spectra using the appropriate weights to
match the slit illumination profile of a
galaxy. Small displacements away from the slit center cause small shifts of the spectrum
along the dispersion
axis. We must correct for this effect, in order to derive the galaxy
dynamics as accurately as possible. For calibration purposes, we
also obtained spectra of the internal wavelength calibration source (wavecals) and the
internal continuum lamp. This contemporaneous flatfield spectrum was obtained 
(while HR~7615 was occulted
by the earth) through the $0\farcs3 \times
0\farcs09$ (length $\times$ width) slit rather than the long-slit to simulate the illumination pattern of a point source
at the detector plane,
because the internal continuum lamp illuminates the slit plane uniformly.
The purpose of such contemporaneous flatfields is to provide for proper calibration of the internal fringing
that is significant at $\lambda \ge 7500$ \AA
\ (see Goudfrooij, Baum, \& Walsh~1997).

The STIS CCD observations of NGC~1023 were obtained on 1997 November 13.
The slit was aligned at a P.A. of
$93\arcdeg$ (observing constraints prevented the slit from being aligned more closely
with the major axis). Table~1 lists the detailed configuration of STIS during this
observation. We integrated for three HST orbits at the calcium triplet (using the
grating G750M)  
for a total integration time of 7574 sec. Two successive exposures were obtained 
during each orbit to
identify cosmic ray events, and NGC 1023's nucleus was shifted by 4 pixels along the slit
before resuming integration in the next orbit. Wavecals
were
interspersed among the galaxy exposures to allow for wavelength calibration with correction of
thermal drifts during
data reduction. Contemporaneous flatfields through the same long-slit 
were obtained while NGC~1023 was
occulted by the earth.

The reduction of STIS spectroscopic data covering the calcium triplet is not
straightforward. The reduced data for this STIS mode provided by the STScI
pipeline are not optimal. Their technique for rejection of hot pixel
events (see Kimble et al.~1998) relies on daily monitoring of the dark image.
However, we found that
hot pixels stronger than 9$\sigma$ above the background do not subtract very
well, presumably because such hot pixels vary in intensity 
on timescales of less than one day. Also, the STScI pipeline uses a library flatfield
rather than the contemporaneous flatfield obtained during the observations, and their 
geometric
rectification is not sufficient for our purpose. With these limitations noted, we
found that the basic reduction steps (bias subtraction, dark subtraction, and
combining the sub-exposures to reject cosmic ray events) of the CALSTIS task in the 
STIS package of
STSDAS could be made to work, if the dark calibration image employed 
was first median filtered to reject
hot pixels $> 9\sigma$ above the background. These strong hot pixels were
cleaned out later (see below). For the remainder of the reduction, we utilized
standard procedures in IRAF for long-slit spectroscopy.   
For each science spectrum, 
we constructed a fringe flat from its associated contemporaneous flatfield, 
using a third-order polynomial to normalize
the flatfield and remove the color of the lamp. We tested the alignment of the
fringe pattern between the contemporaneous flatfield images and the data 
by shifting the phase and amplitudes of the contemporaneous fringe flats
by arbitrary factors and then dividing these into the data. Based on the
residuals seen in the continuum (compared to our KPNO spectra), 
we found that the original
unshifted and unscaled fringe flats were the best flatfield for the data. 
The remaining hot pixels were then cleaned out of each image in the following way.
We divided each image by a heavily median filtered version
of itself, to emphasize hot pixels near the galaxy nucleus 
where the intensity profile
along the slit is steep. The resulting images were then cleaned of hot
pixels using a local median
filter that replaced any pixels deviant from its neighbors by more than $3\sigma$ with
the median value of its neighbors. Neighboring pixels were defined either
by a $5 \times 1$
pixel box near the galaxy center with the long dimension in the dispersion direction,
or by a $5 \times 5$ box away from the galaxy center. These cleaned 
images were then multiplied by the original median filtered image to restore the 
intensity
profile along the slit. We tested this cleaning process extensively to ensure that
the data were not significantly compromised.  
Wavelength solutions and corrections for the geometric 
distortions were determined from the
wavecals and traces of the position of the continuum
peak across the CCD. Although the position of the continuum peak near the central row of the CCD
varies at a uniform rate as
a function of CCD column number, the geometric distortion is not a simple rotation because
the positions of the emission lines in the wavecals do not vary at a uniform rate as a
function of CCD row number. The data were then resampled only once to register the 
galaxy center
to the same row in each image and to bin the dispersion axis on a $\log \lambda$
scale. After this geometric correction had been applied, the rms residual about a 
perfect correction was 0.05 pixel and
demonstrated no systematic behavior with CCD pixel coordinates.
We combined the individual galaxy exposures by taking the average (weighted by
exposure time).
Finally, we applied STScI's pipeline sensitivity curve to flux-calibrate the data,
since this sensitivity curve has a low order structure that, if not
removed, would adversely affect our ability to measure the wings of very broad
galaxy absorption line 
profiles. The HR~7615 spectra have at least 44,000 $e^-$ pixel$^{-1}$ 
(depending on the star's 
position with respect to the slit 
center). Since our flatfield image has only $\approx 11,000$ $e^-$ pixel$^{-1}$, the $S/N$ in
the HR~7615 spectra is limited to 150/\AA. 
For NGC~1023, the $S/N$ is $40/$\AA \ in the row centered on the nucleus.  
As mentioned above, we combined the individual
spectra of HR~7615 to match the slit illumination pattern of NGC~1023 along the 
dispersion axis.

\section{Measurement of the Stellar Kinematics}

In this section, we briefly review the method we utilized to measure the
stellar kinematics from the spectra. Next we examine possible effects of the
STIS/CCD PSF on the stellar kinematics, and then verify that those effects are
represented adequately during construction of galaxy dynamical models in \S 4. 

\subsection{The Fourier Correlation Quotient method} 

Using the Fourier correlation quotient (FCQ) method (Bender~1990) on the ground-based
and STIS spectra,
we measure the line of sight velocity distribution (LOSVD)
$f(v)$ for each extracted spectrum along the slit (where the extracted apertures
were defined to match the binning scheme of the three-integral models
discussed in \S 4.2). Those models require as input
the non-parametric LOSVD as a function of position in the
galaxy. FCQ also determines the moments of the LOSVD which are needed for the
two-integral models constructed in \S 4.1.   
These moments are determined by fitting the LOSVD
with a Gaussian (with radial velocity $V$ and
velocity dispersion $\sigma$) plus third and fourth order
Gauss-Hermite polynomials $H_3$ and $H_4$ (van der Marel \& Franx~1993;
Gerhard~1993; Bender, Saglia, \& Gerhard~1994), i.e.,

\begin{equation}
f(v) = {\gamma \over \sqrt{2 \pi} \sigma} e^{-(v-V)^2 \over 2 \sigma^2}
\left [ 1 + h_3 H_3 \left ({v-V \over \sigma} \right ) + h_4 H_4 \left ({v-V \over \sigma}
\right )
\right ] .
\end{equation}

\noindent
where $\gamma$ is the strength of the galaxy absorption line relative to that of
the template star.
The coefficients $h_3$ and $h_4$, respectively, quantify the lowest order asymmetric
and symmetric deviations of the LOSVD from a Gaussian. If $h_3 > 0$, then
the blue side of the line profile is steeper than the red side, while
the opposite is the case if $h_3 < 0$. If $h_4 > 0$,
then the line profile is more triangular than a Gaussian, i.e., it is more
strongly peaked and has wings that are more extended. If $h_4 < 0$,
then the line profile is more rectangular than a Gaussian, i.e., it has a flatter
top and less extended wings. Non-zero values of $h_3$ can arise when the
galaxy has two or more velocity components, or if the radial rotation gradient is
steep such that both low and high velocities are seen in projection along the
line of sight. Non-zero values of $h_4$ indicate that the
galaxy's velocity distribution function is anisotropic. Fig.~2 shows the FCQ results
for the nuclear row of the STIS spectrum of NGC 1023. We determine the errors in
the LOSVDs by Monte Carlo simulation. This involves convolving the template spectrum with
the FCQ estimate of the LOSVD, and (given the $S/N$ in an extracted galaxy spectrum)
finding thirty different noise realizations of a synthetic galaxy spectrum.
For simplicity, we assume that the noise distribution in amplitude is Gaussian, which is
nearly identical to a Poisson distribution for these high count levels.
The synthetic galaxy and template spectra are then input into FCQ, providing a
sample of the error distribution in the LOSVD. 

Fig.~3 shows the Gauss-Hermite
moments measured from the STIS and ground-based spectra as a function of distance
from the nucleus. A preliminary examination of these data show that $\sigma$
increases significantly toward the nucleus (where $\sigma = 295 \pm 30$ km s$^{-1}$),
which could indicate the presence of a nuclear BH. 
In \S 4, the data in Fig.~3 will be compared
to
dynamical models. In the meantime, next we examine the effects of the STIS/CCD PSF on
the data. 

\subsection{Effects of the STIS/CCD PSF}

During our initial application of FCQ to the STIS spectrum of
NGC~1023, we became aware that the STIS PSF might be asymmetric
at $\sim 8500$~\AA. Given this
galaxy's
steep power law surface brightness profile (Fig.~1), we were concerned
that the off-nuclear spectra might
be distorted by the nuclear spectrum in a way that is asymmetric along the slit.
During the construction of galaxy dynamical models (\S 4), we convolve the dynamical
models with
a modeled PSF before comparison with the observations. This step is intended to
correct for the effects of the PSF, but this model PSF must be circularly
symmetric. It would not be straightforward at all to allow for an asymmetric PSF
when constructing dynamical models. Besides, we can measure the STIS PSF at this wavelength
only along the slit (which we describe below), so we have no knowledge about its 
two-dimensional structure. 
[Although STIS has an imaging mode, there are no filters with 
sufficiently narrow bandpass in the near-infrared to isolate the PSF asymmetry which
is caused by diffraction effects.]  

We measure the STIS PSF from the spectrum of the template star HR~7615. The main
complication is that at
$\sim 8500$~\AA, the FWHM of the PSF is approximately 1.5 pixels, so
the PSF is undersampled. Consequently, a simple cut along the
spatial axis of a point source spectrum does not provide an adequate description
of the PSF because the profile is adversely affected by aliasing. One way to recover 
an oversampled measurement of the PSF from
such data would be to dither a point source
along the slit by fractional pixels between exposures. To find the PSF, the profiles 
along the slit
would then be combined by interleaving the pixels.

Even though during our observation HR~7615 was not dithered along the spatial axis of 
the slit by
moving the telescope, the geometric distortions in the STIS/CCD camera are such that
the position of the stellar peak varies slowly as a function of distance along the
dispersion axis in the raw data. The amplitude of this geometric distortion is 
$\approx 6$ rows
across the 1024 columns along the dispersion axis. The rate that the position
of the stellar peak changes per CCD column is approximately constant as a function of CCD 
column number (see \S 2), so the spatial profile of the source
is effectively dithered along the slit by $\approx 0.006$ pixels per column. Using this
effect to our advantage, we reconstruct an oversampled measurement of the PSF using the
spectrum of HR~7615 with the star centered in the slit. For the present purpose, the
input spectrum is from an intermediate step during data reduction (just prior to
geometric correction).
The PSF is determined as a function of position along the dispersion axis in 
spectral bins 9 columns
wide. Such binning is necessary to obtain a PSF with very high $S/N$ for
accurate centering. The relative 
geometric distortion across a 9 pixel bin is $\sim 0.054$ pixels which is negligible
compared to the PSF FWHM ($\approx 1.5$ pixels). The PSF from each of the 
bins is prepared to be
interleaved with the PSF from the other bins by normalizing each by the
total flux (within a distance of 10 pixels of the peak) along the slit and shifting the pixel 
coordinates so that the centroid of
the PSF in each bin is identical. The pixel intensity values are not altered in any way during
this process. We also measured the PSF from the spectrum of
the K3III star HR~260 obtained by Bower et al.~(2000) on 1998 March 15,
using the same instrumental setup as our HR~7615 visit. These reconstructed PSFs are 
shown in Fig.~4. The PSFs derived from separate stars agree extremely well: there was
no variation in the PSF at this epoch on a timescale of 5 months.

The PSF in Fig.~4 has FWHM = 1.58 pixels ($0\farcs079$) and has 
broad asymmetric wings, which peak at $R = \pm$ 2 pixels.
These wings represent the first Airy ring
in the PSF, and are asymmetric probably because the two-element corrector optics in
STIS might not be perfectly aligned for this wavelength
(C.~W.~Bowers, private communication). The purpose of
these corrector
optics is to correct for the HST primary mirror's spherical aberration and 
for the astigmatism at the
STIS field of view in HST's aperture plane (Kimble et al.~1998; Woodgate et al.~1998).
 
Could the asymmetric 
wings on the PSF
significantly distort our measurements of the stellar kinematics? 
To test this possibility, we construct a simulated spectrum of NGC~1023 from the
spectrum of HR~7615 using the empirical fit to NGC~1023's surface brightness profile
$\mu_V(R)$
(dashed line in Fig.~1) and a simple empirical model
of the stellar kinematics.  
In this model, $V(r)$ and $\sigma(r)$ are set to simple 
analytical
forms (representative of the $V(r)$ and $\sigma(r)$ profiles determined from the
observed galaxy spectrum) and $h_3$ and $h_4$ are set to zero at all radii.  
This model galaxy spectrum is
then input into a STIS simulator that: (1) applies the spectrograph's
sensitivity curve and the integration time to convert from flux
units to detector counts, (2) convolves the spectrum with an input PSF, and 
(3) adds Gaussian 
noise to simulate the $S/N$ in the
real observation. For each simulation, four spectra are saved to cover all
possibilities with steps (2) and (3) independently turned on or off. 
We generate two
sets of simulated spectra: one for the observed PSF from Fig.~4, and one for the model
PSF used during
modeling in \S 4. This model PSF was constructed from the observed PSF by folding it about
$R=0$ and then determining the average within radial bins of 0.1 pixel. The purpose of
these simulations is
to examine the effects of the observed PSF and to check that the model PSF faithfully
reproduces any such effects within the errors.

We then input all simulated spectra into FCQ to examine the effects of the PSF and noise,
yielding the results in Fig.~5. Fig.~5a shows that the most significant effect
of the PSF is to smear out the $\sigma$ profile, as expected. The most important result,
however, is seen in Fig.~5b. The kinematic measurements from
the simulated spectra using the observed PSF or our model PSF are not statistically 
different,
even when
noise is not added to the spectrum. This demonstrates that the model
PSF reproduces the effects of the observed PSF, and justifies our approximation that the PSF
is symmetric.  

\section{Dynamical Modeling}

In this section, we construct galaxy dynamical models from the
observed surface photometry and stellar kinematics, to determine if the observations
require the presence of a dark compact object. We utilize the independent
modeling algorithms of Magorrian et al.~(1998) and Gebhardt et al.~(2000c). Both 
of these references
should be consulted for a general discussion of these techniques. In this paper
we provide only a brief summary and the details of the models constructed for
NGC~1023.
Both algorithms assume that the galaxy mass distribution is axisymmetric
with constant inclination ($i$, with respect to the plane of the sky). 
That assumption is justified by
the behavior of PA and ellipticity as a function of radius at $R < 10''$ (Fig.~1),
which are nearly constant with radius at values of $83\arcdeg$ and $0.28$, respectively.
A non-axisymmetric density distribution (such as a triaxial one) should exhibit significant
radial variations in one or both of these quantities. The two modeling algorithms
differ dramatically in the way that the internal dynamics are modeled.
Magorrian et al.'s
algorithm assumes that the velocity distribution function is a function of only
the orbital energy $E$ and angular momentum $L_z$ about the polar axis with
an isotropic velocity dispersion tensor. It then
solves the Jeans equations to
construct a model.
This process is straightforward for a given galaxy. Gebhardt et al.'s algorithm assumes three
integrals of motion, $E$, $L_z$, and $I_3$ (see Binney \& Tremaine 1987). It utilizes 
Schwarzschild's (1979) orbit-based method constrained
by Richstone \& Tremaine's (1988) maximum entropy technique. Consequently, the three-integral
models are more computationally intensive to construct but allow for a fully
general axisymmetric model. 
This section is subdivided to discuss the results of each modeling algorithm when applied
to NGC~1023.

\subsection{Two-integral Models}

Magorrian et al.'s (1998) method has four steps, which stated briefly are: (1)
assuming a value $i$,
identify a smooth luminosity density $\nu(R,z)$ that projects to an acceptable fit
of the observed surface brightness profile $\mu(R)$; (2) calcuate the gravitational
potential using assumed values for the mass to light ratio ($M/L$) and nuclear black hole
mass ($M_{BH}$); (3) solve the Jeans equations for the second-order moments 
($\nu \bar{v_{\phi}^2}$ and $\nu \bar{v_R^2} \equiv \nu \bar{v_z^2}$) of the distribution
function; and (4) project the luminosity-weighted zeroth- and second-order moments
of the line of sight velocity onto the sky, convolve with the instrumental PSF, and
average over the same bins as in the observations. Thus, the free parameters for any model
for our observations are $i$, $M/L$, and $M_{BH}$. 

The PSF is an important ingredient in the modeling. 
Since we have kinematic observations from HST and
the ground and the PSFs are very different, we apply the appropriate PSF 
to model the different data points. However, we do not model more than one observation at a given
position in the galaxy, so we use STIS data within $4''$ of the nucleus along the major
axis and ground-based data at all other positions.  
The PSF for ground-based observations is assumed to be Gaussian
with the appropriate FWHM as described in \S 2. For the STIS data,
the model PSF described in
\S 3.2 is adopted.

Our galaxy models assume that the inclination is constant within the galaxy.
Since NGC~1023 is 
classified as an SB0, we are assuming that the inclination of the
bulge and disk are identical. The ellipticity
at radii beyond the effective radius of the bulge ($R_e = 36''$, Barbon \& Capaccioli 1975)
is $\approx 0.60$ and provides a constraint on the inclination. If the vertical
thickness of the outer stellar 
disk is negligible, then the disk would have $i = 66\arcdeg$. This is a lower
limit on the true inclination, because the thickness of the outer disk actually might 
not be
negligible. Consequently, we will construct models only 
with $i \geq 66\arcdeg$.    
Fig.~6 shows the kinematics (i.e., the second-order moment $\mu = [V^2 + \sigma^2]^{1/2}$
as a function of radius)
of the best-fitting two-integral models for the major and minor
axes for $i=90\arcdeg$ and $i=66\arcdeg$. We do not include the observational data
along the intermediate axis in these model fits because doing so caused
poor model fits for all three axes. This motivates our construction of
three-integral models in \S 4.2. For evaluating the goodness-of-fit of
the two-integral models, the figure
of merit is $\chi^2 = \sum ((\mu_i - \hat{\mu}_i)^2/\sigma_i^2)$, where $\mu$ and
$\hat{\mu}$ are the observed and modeled second-order moments, and $\sigma_i$ is the
error in $\mu_i$ describing the 68\% confidence band.
We find that $\chi^2$ is insensitive to reasonable
values of $i$ (see above), so we average the best-fitting parameter values over this
range in $i$. 
This yields
$M_{BH} = (6.0 \pm 1.4) \times 10^7 \ M_{\odot}$ [which corresponds to
$M_{BH}/M_{bulge} = (1.1 \pm 0.2) \times 10^{-3}$], and $M/L_V = 5.38 \pm 0.08$.
Fig.~7 shows the posterior distribution $\Pr(M/L_V,M_{BH}\mid D$) which is the
likelihood based on $\chi^2$ of certain values of $M/L_V (\equiv \Upsilon)$ and $M_{BH}$ given the
photometric and kinematic data $D$. See Magorrian et al.~(1998) for a detailed definition
of $\Pr$. 
These model results provide a useful
estimate of the area of parameter space that we will explore more intensively with the
three-integral models.

\subsection{Three-integral Models}

Gebhardt et al.'s (2000c) method has three steps, which stated briefly are:
(1) assuming a value for $i$ and that the mass distribution's ellipticity is
constant with radius, deproject the observed surface brightness profile
(as in Gebhardt et al.~1996)
to estimate the luminosity density (see Fig.~8); (2) working in units of $M/L$ to
minimize the computation time required, combine the luminosity density profile
with an assumed value for central black hole mass $M_{BH}$ to determine the library of 
representative stellar orbits
that would be possible in that gravitational potential; and (3) for an adopted value
of $M/L$, superimpose the orbits from the library (under constraint by maximum
entropy) to fit the observed LOSVD as a function of position in the
galaxy. These three steps
are repeated as necessary to vary $i$, $M/L$, and $M_{BH}$.

The models are binned spatially (as projected onto the sky)
and in velocity space (since we are modeling the
LOSVD non-parametrically). The spatial bins are in radius $r$ 
(from the galaxy center)
and $\theta$ (the polar angle measured from the major axis). 
During step (2), we use 80 radial bins, 20 angular bins,
and 15 velocity bins, which are approximately logarithmic in $r$, linear in 
$\sin\theta$, and linear in 
velocity. The bins in $r$ and 
$\theta$ cover $0\farcs0025$ to $317''$,
and $0\arcdeg$ to $90\arcdeg$, respectively. At small radius where the bins converge, 
some bins are combined with neighboring bins as needed to mimic the STIS
slit width. The bins in velocity are
defined in units of $\sqrt{M/L}$ (in solar units) and cover $-337$ to $+337$. The coverage and
size of the velocity bins are chosen to include the wings of the LOSVD for the
maximum velocity dispersion observed, and to match approximately the instrumental 
resolution. During step (2), we follow $\sim 3300$ orbits per orbital direction (i.e.,
prograde and retrograde), for a total of $\sim 6600$ orbits. During step (3) when comparing the 
model to the 
data, the
$r$ and $\theta$ bins are compressed
from $80 \times 20$ to $20 \times 5$ for
computational expediency.

The effect of the instrumental PSF is incorporated during step (3). As discussed in \S 4.1, 
we use different PSFs 
for the HST and ground-based data as appropriate, and we fit only the STIS data at
radius $< 4''$ along the major axis and only the ground-based data at other positions
for each of the three PA's we observed.

The advantage of our model setup is that almost all model bins correspond exactly
to an observational bin, so comparing the model to the data requires almost no
interpolation of the model. During the model construction, we match  
the light in the spatial
bins defined
above and fit the LOSVDs in the 37 extracted aperture positions where we have data.
Therefore, the maximum number of observables is 555 (i.e., 37 positions times
15 velocity bins). However, the actual number of independent observables is smaller 
than
this because FCQ derives the LOSVD with the assistance of an optimal filter
in Fourier space (see Bender 1990), whose effect depends on the significance of
noise across the cross correlation function (CCF)
between galaxy and template spectra. The filter is more
significant for the wings of the CCF where it is noisier relative to
the peak. Calculating the actual number of observables would be complicated.
Instead, we assume that half of the velocity bins in the LOSVD (i.e., those
near the peak) are uncorrelated, and the other half are correlated with their
adjacent bins. Under that assumption, the number of independent observables would be
$\sim 340$.   
Our relative figure of merit is $\chi^2 = \sum ((y_i - Y_i)^2/\sigma_i^2)$, 
which
is the sum over all 555 observables of the deviation between the observational
data points $y_i$ and their model fits $Y_i$ given the observational errors $\sigma_i$.
Fig.~9 shows two example LOSVD model fits to the nuclear LOSVD for models
with $M_{BH} = 5.6 \times 10^7 \ M_{\odot}$ and zero
(which are discussed below). Since the
actual number of degrees of freedom cannot be determined very well, we do not
attempt to determine the goodness-of-fit for a given model from its value of $\chi^2$. 
Instead, we
follow the change in $\chi^2$ as a function of the three model variables ($M_{BH}$,
$i$, and $M/L$), and for each model calculate $\Delta \chi^2$ with respect to the
minimum $\chi^2$ given by the best 
fitting model. 

For the best model, the total $\chi^2$ is $274$. If the 
number of degrees of freedom is 340, then the reduced
$\chi^2$ would be $\sim 0.8$. If the reduced $\chi^2$ should be unity, then
we have overestimated either the number of degrees of freedom or
the error bars on the LOSVDs.  
Since this small offset in reduced $\chi^2$ is within our uncertainty of the 
estimated number of
degrees of freedom, we conservatively accept $\Delta \chi^2 = 1$ to define the
68\% confidence interval in parameter space.

Our strategy for covering parameter space is to
first run models for a low resolution grid that covers 
$M_{BH}$ from
zero to $2 \times 10^8 \ M_{\odot}$,
$i$ of $90\arcdeg$ to $70\arcdeg$, and $M/L_V$ of $5.0 - 7.4$ in solar units.
Based on that grid, we find a preliminary estimate for the location of the $\chi^2$
minimum. Then we run models for a higher resolution grid concentrated around
the apparent minimum.
Although the two-integral models in \S 4.1 are
insensitive to the value of $i$, the three-integral models are very sensitive to $i$.
The best fitting models have $i=90\arcdeg$. In fact, for models with
$i \leq 80\arcdeg$, $\chi^2$ increases
to at least 400. Since the fits of the edge-on models are vastly superior, 
we will not discuss the $i \neq 90\arcdeg$ models in detail any further.
However, the most likely reason for this difference is our exclusion of a
dark halo, since that limits our ability to model successfully the kinematics
at large radii. To find the confidence
band in
parameter space, very good coverage in parameter space is beneficial. However,
the models are computationally intensive. Since the surface of $\chi^2$ is 
relatively smooth, we utilize a bicubic spline interpolation of the values of 
$\chi^2(M_{BH},M/L_V)$ to estimate the value of $\chi^2$ where we do not have models.
Fig.~10 shows these contours of $\Delta \chi^2$
superimposed on a grid showing the positions in
parameter space of the useful
edge-on models.  
The best model has $M_{BH} = 4.2 \times 10^7 \ M_{\odot}$ and 
$M/L_V = 5.57$. Fig.~11 shows $\chi^2(M_{BH} \mid M/L_V=5.6)$,
a one-dimensional projection of $\chi^2$ 
near the minimum. 
Fig.~11 indicates that $\chi^2$
near the minimum 
changes slowly as a function of BH mass, since it is constrained only by
the central few LOSVD measurements. The models
are very sensitive to $M/L_V$ because the total mass is well constrained.
The three-integral models fit the entire LOSVD at each
point along the slit, whereas the two-integral models fit only the 
profile of the second moment.  
Given the model assumptions, the range of values 
in the formal 68\% confidence region is $M_{BH}$ of
$3.5 \times 10^7 - 4.3 \times 10^7 \ M_{\odot}$ and $M/L_V$ of $5.54 - 5.58$.

The three-integral models fit the observed LOSVD at each observed position in
the galaxy, yet we also determine the model Gauss-Hermite moments by fitting equation 
(2) to the model LOSVDs. This provides for a comparison between the observations and models
where the effects of a BH are more straightforward to visualize (relative to comparisons
between the observed and model LOSVDs, e.g., Fig.~9). In Fig.~3, we show the 
Gauss-Hermite moments for three of the models 
(including a model near the best fit, and models without a BH and where the BH
is much too massive), superimposed on the data. Although this comparison in Fig.~3 is
instructive, this is not the optimal way of comparing models to the data, since 
$\chi^2$ is determined by the fit to the LOSVD, not to its moments. Also, since
we do not include a dark halo component, we cannot expect to fit the kinematic data
at radii $\gtrsim 100''$ (5 kpc). Nevertheless,
it is now straightforward to see the reason that models without a BH fail. Along
the major axis, the rotation is rapid ($V \approx 70$ km s$^{-1}$ at $R = 0\farcs1$),
while $\sigma$ is increasing toward the nucleus. For models without a BH, fitting the
rapid rotation requires populating more tangential orbits, yet overpopulating
a single class of orbits restrains $\sigma$
from increasing toward the nucleus. Alternatively, the model with
$M_{BH} = 2.2 \times 10^8 \ M_{\odot}$ yields a $\sigma$ profile that is too
steep toward the nucleus. The model with
$M_{BH} = 5.6 \times 10^7 \ M_{\odot}$ achieves a compromise between these two extremes.  

To summarize this section, we have constructed both two-integral and three-integral
axisymmetric dynamical models. Both methods point strongly to the presence of a nuclear
dark compact object (which presumably is a supermassive black hole).
The best-fitting values of BH mass and $M/L_V$ from the two methods are not quite 
in formal
agreement. The BH and $M/L_V$ results from the two-integral models differ from those
from the three-integral models by 53\% and 3\%, respectively.

\section{The Internal Dynamics of NGC~1023}

In this section, we consider the implications for the internal structure of
NGC~1023 that can be drawn from our data and modeling. In particular, 
Faber et al.~(1997) suggested that NGC~1023 probably has a nuclear stellar disk,
based on its surface photometry. Is this suggestion consistent with the dynamical
structure seen in our
best-fit model?       

Fig.~12 shows the internal dynamics of the best-fitting model. The three
components of the internal velocity dispersion
in this model 
are nearly isotropic at $r > 0\farcs4$ ($19.6$ pc). Within this radius as one approaches 
the
nucleus, the anisotropy changes from radial anisotropy to tangential
anisotropy in the central bin. This is indicated by $\sigma_r/\sigma_t < 1$ and
$\sigma_{\phi}$ is much greater than $\sigma_r$ or $\sigma_{\theta}$ at the
nucleus. 
This tangential anisotropy at the nucleus also suggests the presence
of a nuclear stellar disk. 
The low value of v$_{\phi}$ at the nucleus implies that the stars in this
nuclear disk are
nearly evenly divided between prograde and retrograde orbits.

The internal dynamics point to tangential anisotropy at the nucleus, which
is normally associated with negative values of $h_4$ when projected onto the
sky. However, the nuclear value of $h_4$ is significantly
positive (Fig.~3), and the nuclear LOSVD has wings at velocities $\geq 
600$ km~s$^{-1}$ from systemic (Fig.~9). 
These strong tails likely indicate that, in addition to the population of stars in
the nuclear disk, there are stars
very close to the central BH with very large velocities 
(either radial or
tangential) that are enhancing the wings of the nuclear LOSVD. This is consistent
with the existence
of a nuclear star cluster (\S 2.1) sufficiently bright to
influence the central LOSVD.

Gebhardt et al.~(2000a) and Ferrarese \& Merritt (2000) have shown that, for
galaxies with the best dynamical evidence for a BH, the mass of the BH is strongly
correlated to the velocity dispersion $\sigma_e$ within the half-light radius $R_e$.
NGC~1023, with a BH mass of $3.9 \times 10^7 \ M_{\odot}$ and
$\sigma_e = 205$ km~s$^{-1}$, lies slightly below (with $2\sigma$ confidence)
the ridge line of this relation. Although this offset is not significant,
determining BH masses with comparable accuracy as NGC~1023 will probe the
possible intrinsic scatter in this correlation. Given the observed
high velocity population of stars in the nucleus of NGC~1023, this
galaxy is important since
it may be one of the few galaxies where we actually observe
the direct influence of the BH on the kinematics.
Isolating this high velocity nuclear population would provide even stronger
constraints on the nuclear mass distribution.


\acknowledgments

We acknowledge several very useful suggestions from John Kormendy
and Gregory Bothun.
This work was supported by NASA Guaranteed Time Observer funding to the STIS 
Science Team.


\clearpage

\par\noindent
{\bf References}

\par\noindent
Barbon, R., \& Capaccioli, M.~1975, \aap, 42, 103\\
Bender, R.~1990, \aap, 229, 441\\
Bender, R., Saglia, R.~P., \& Gerhard, O.~E.~1994, \mnras, 269, 785\\
Binney, J., \& Tremaine, S.~1987, Galactic Dynamics (Princeton: Princeton
University Press)\\
Bower, G.~A., et al.~2000, in preparation\\
de Vaucouleurs, G., et al.~1991, Third Reference Catalog of Bright Galaxies
(Springer: New \\
\indent York)\\ 
Faber, S.~M., et al.~1997, \aj, 114, 1771\\
Ferrarese, L., \& Merritt, D.~2000, \apj, 539, L9\\
Gebhardt, K., et al.~1996, \aj, 112, 105\\
Gebhardt, K., et al.~2000a, \apj, 539, L13\\
Gebhardt, K., et al.~2000b, in preparation\\
Gebhardt, K., et al.~2000c, \aj, 119, 1157\\
Gerhard, O.~E.~1993, \mnras, 265, 213\\
Goudfrooij, P., Baum, S.~A., \& Walsh, J.~R.~1997, in The 1997 HST
Calibration Workshop,\\
\indent eds.~S.~Casertano, et al.~(Baltimore: STScI), 100\\
Kimble, R.~A., et al.~1998, \apj, 492, L83\\
Kormendy, J.~2000, in Galaxy Disks and Disk Galaxies, eds.~J.~G.~Funes, S.~J., 
\& E.~M.\\
\indent Corsini (San Francisco: ASP), in press (astro-ph/0007401)\\
Kormendy, J., \& Bender, R.~1999, \apj, 522, 772\\
Kormendy, J., \& Richstone, D.~1995, \araa, 33, 581\\
Lauer, T.~R., et al.~1995, \aj, 110, 2622\\
Lauer, T.~R., et al.~2000, in preparation\\
Lucy, L.~B.~1974, \aj, 79, 745\\
Magorrian, J., et al.~1998, \aj, 115, 2285\\
Maoz, E.~1998, \apj, 494, L181\\
Richardson, W.~H.~1972, JOSA, 62, 55\\
Richstone, D., \& Tremaine, S.~1988, \apj, 327, 82\\ 
Richstone, D., et al.~1998, \nat, 395, A14\\
Schlegel, D.~J., Finkbeiner, D.~P., \& Davis, M.~1998, \apj, 500, 525\\ 
Schmidt, M., Schneider, D.~P., \& Gunn, J.~E.~1995, \aj, 110, 68\\
Schwarzschild, M.~1979, \apj, 232, 236\\
Simien, F., \& deVaucouleurs, G.~1986, \apj, 302, 564\\
van der Marel, R.~P., \& Franx, M.~1993, \apj, 407, 525\\
Woodgate, B.~E., et al.~1998, \pasp, 110, 1183\\


\figcaption
{Major axis surface photometry of NGC~1023, including surface brightness
$\mu_V$, position angle PA, ellipticity $\epsilon$, and ($V-I$). The
fits to the surface brightness profile
include a Nuker law (solid line) and a Nuker law plus model for the nuclear
stellar cluster (dashed line). \label{fig1}}

\figcaption{Results from the FCQ method
for the nuclear row of the STIS spectrum of NGC 1023. The top panel shows the galaxy
spectrum (black) and the template spectrum (red) after it has been
convolved with the parametric LOSVD for this row. Both spectra have been continuum 
subtracted and have been tapered
at the ends using a cosine bell. The lower panel shows the non-parametric (black) and 
parametric (red)
LOSVDs. The dashed curves represent the 68\% confidence interval for the non-parametric
LOSVD. The Gauss-Hermite moments for this fit are $V = 595 \pm 18$ km s$^{-1}$,
$\sigma = 295 \pm 30$ km s$^{-1}$, $h_3 = -0.021 \pm 0.056$, and
$h_4 = 0.162 \pm 0.056$. \label{fig2}}

\figcaption{Gauss-Hermite moments measured from the STIS and ground-based spectra
utilizing the Fourier Correlation Quotient method. Each column of panels shows the
four Gauss-Hermite moments for a given slit position angle that we observed.
The curves show three of the edge-on axisymmetric models that we construct in \S 4.2.
All three have $M/L_V = 5.6$, and include the best-fitting BH of $5.6 \times 
10^7 \ M_{\odot}$ (solid line), no BH (dashed blue), and a BH of 
$2.2 \times 10^8 \ M_{\odot}$ (dotted brown). 
\label{fig3}}

\figcaption{Reconstructed STIS/CCD PSFs using the spectra of the
template stars HR~7615 (blue points) and HR~260 (red points). The agreement between the
two PSF measurements is so good that the two are almost indistinguishable.
The sign convention on the abscissa 
is defined
such that increasing $R$ corresponds to increasing row number along the
spatial axis on the CCD. \label{fig4}}

\figcaption{Stellar kinematics measured from simulated STIS spectra
of NGC~1023 to investigate possible effects of the asymmetric PSF.
Solid lines represent the simple empirical kinematical model from which the
simulated spectra were constructed. (a) Simulations without noise added to the
spectra. (b) Simulations that include noise. The red open circles show the measurements 
from the simulated spectrum before convolution with the PSF, and the green crosses
and blue stars represent, respectively, the measurements from the spectrum convolved 
with the PSF from Fig.~4
(i.e., the actual PSF) or the model PSF (see text). The error bars in (a) are no
larger than the size of the points. The most important result is that the green crosses
and blue stars agree, which justifies our approximation in \S 4 that the PSF is symmetric.
\label{fig5}}

\figcaption{Kinematic profiles of the best-fitting two-integral models along the
(a) major and (b) minor axes of NGC~1023. The plots show the second moment
$(V^2 + \sigma^2)^{1/2}$ (in units of km s$^{-1}$)
versus distance $R$ from the center of the galaxy.
The points are the observational data. The curves show the best-fitting models
for $i=90\arcdeg$ and $i=66\arcdeg$. The latter model lies above the
former. \label{fig6}}

\figcaption{Posterior distribution $\Pr(\Upsilon,M_{BH}\mid D$) for the
two-integral models of NGC~1023 (note that $\Upsilon \equiv M/L_V$). 
Successive light contours indicate a factor
of 10 change in $\Pr(\Upsilon,M_{BH}\mid D$). The heavy contours enclose the
68\% and 95\% confidence regions on $M/L_V$ and $M_{BH}$. \label{fig7}}

\figcaption{Radial profiles of surface brightness and luminosity density in V
for NGC~1023 (corrected for Galactic extinction). 
(Left) Open circles represent the HST data from Fig.~1, while closed
squares represent the data from Barbon \& Capaccioli (1975) and the extrapolation beyond
their last data point at $R = 189''$ using an $R^{1/4}$ law with $R_e = 36''$.
The solid line represents a smoothing spline fit to the data. (Right) The inferred
luminosity density for an inclination of $90\arcdeg$. The deprojected density 
distributions for $i = 80\arcdeg$ and $70\arcdeg$ (for which we will also construct
dynamical models) are not shown because their difference from the
$i = 90\arcdeg$ case (1\% and 6\%, respectively) is insignificant on this scale.
\label{fig8}}

\figcaption{Projected LOSVD at the nucleus (open circles) and the model fits to it 
(solid points) for the (a) $M_{BH} = 5.6 \times 10^7 \ M_{\odot}$
model and (b) no BH model. \label{fig9}}

\figcaption{Contours of $\Delta \chi^2$ (with values of 1.0, 3.84, and 6.63).
One-dimensional projections of these contours
correspond to the confidence bands of 68\%, 95\%, and 99\%, on the
black hole mass $M_{BH}$ and mass to light ratio $M/L_V$. The solid points show
the positions in parameter space of the models near the $\chi^2$ minimum. 
\label{fig10}}

\figcaption{One-dimensional projection of $\chi^2(M_{BH},M/L_V)$ onto the axis of
$M_{BH}$ for $M/L_V = 5.6$. 
The points represent the models, and the dashed line is the
projection of the bicubic spline utilized for interpolation. 
Since $M_{BH}$ is on a log scale,
we include the $M_{BH} = 0$ model at $M_{BH} = 10^6 \ M_{\odot}$.
\label{fig11}}

\figcaption{Internal dynamics along the major axis for the best-fitting
model ($M_{BH} = 5.6 \times 10^7 \ M_{\odot}$, $M/L_V = 5.6$,
and $i = 90\arcdeg$). Starting from the top, the separate panels show
the three internal velocity dispersions, the ratio of radial to
tangential velocity dispersion [where $\sigma_t^2 = (\sigma_{\theta}^2
+ \sigma_{\phi}^2)/2$], and v$_{\phi}$ (the mean velocity in the
equatorial plane). \label{fig12}}

\clearpage

\begin{deluxetable}{lll}
\footnotesize
\tablewidth{460pt}
\tablecaption{Instrumentation Details \label{tbl-1}}
\tablehead{
\colhead{Quantity} 
& \colhead{KPNO} & \colhead{HST}}
\startdata
Spectrograph & 
RC & STIS \nl
Observation Dates & 
1997 Feb $15-16$ & 1997 Nov $13$  \nl
Detector & 
Tek $2048 \times 2048$ & SITe $1024 \times 1024$ \nl
Gain (e$^{-}$/ADU) & 
1.9 & 1.0 \nl
Wavelength coverage & 
7490 \AA \ $-$ 9861 \AA &
8272 \AA \ $-$ 8845 \AA \nl
Reciprocal dispersion (\AA \ pixel$^{-1}$) & 
1.14 & 0.56 \nl
Slit width (\arcsec) & 
1 & 0.2 \nl
Comparison line FWHM (pixel) & 
2.0 & 3.1  \nl
R $= \lambda / \Delta \lambda $ & 
3800 & 4930 \nl
Instrumental dispersion ($\sigma_I$) (km s$^{-1}$) & 
60 & 56 \nl
Spatial scale (\arcsec \ pixel$^{-1}$) & 
0.69 & 0.050  \nl
Slit length (\arcmin) & 
5.5 & 0.8 \nl
\enddata
\end{deluxetable}

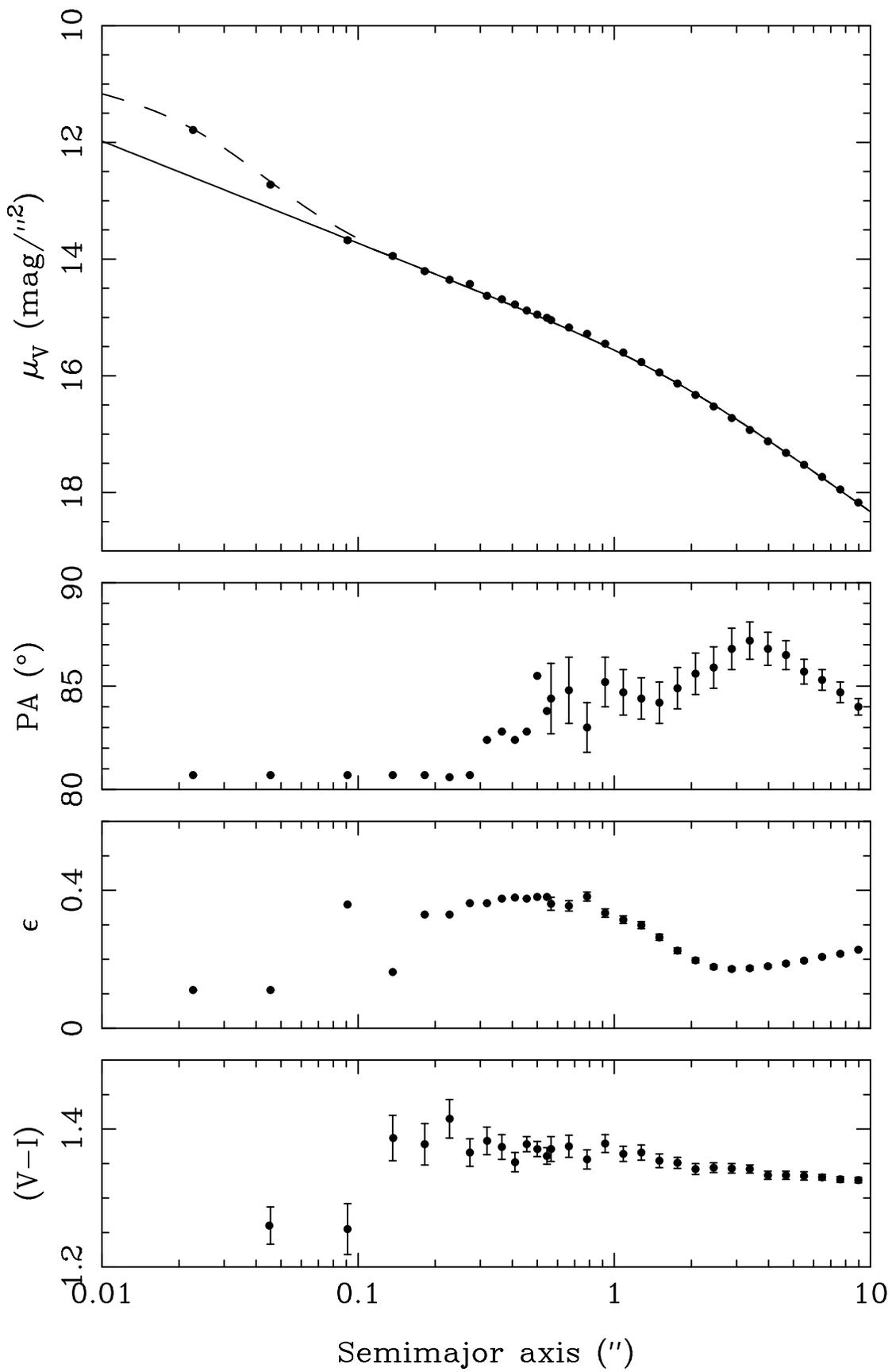
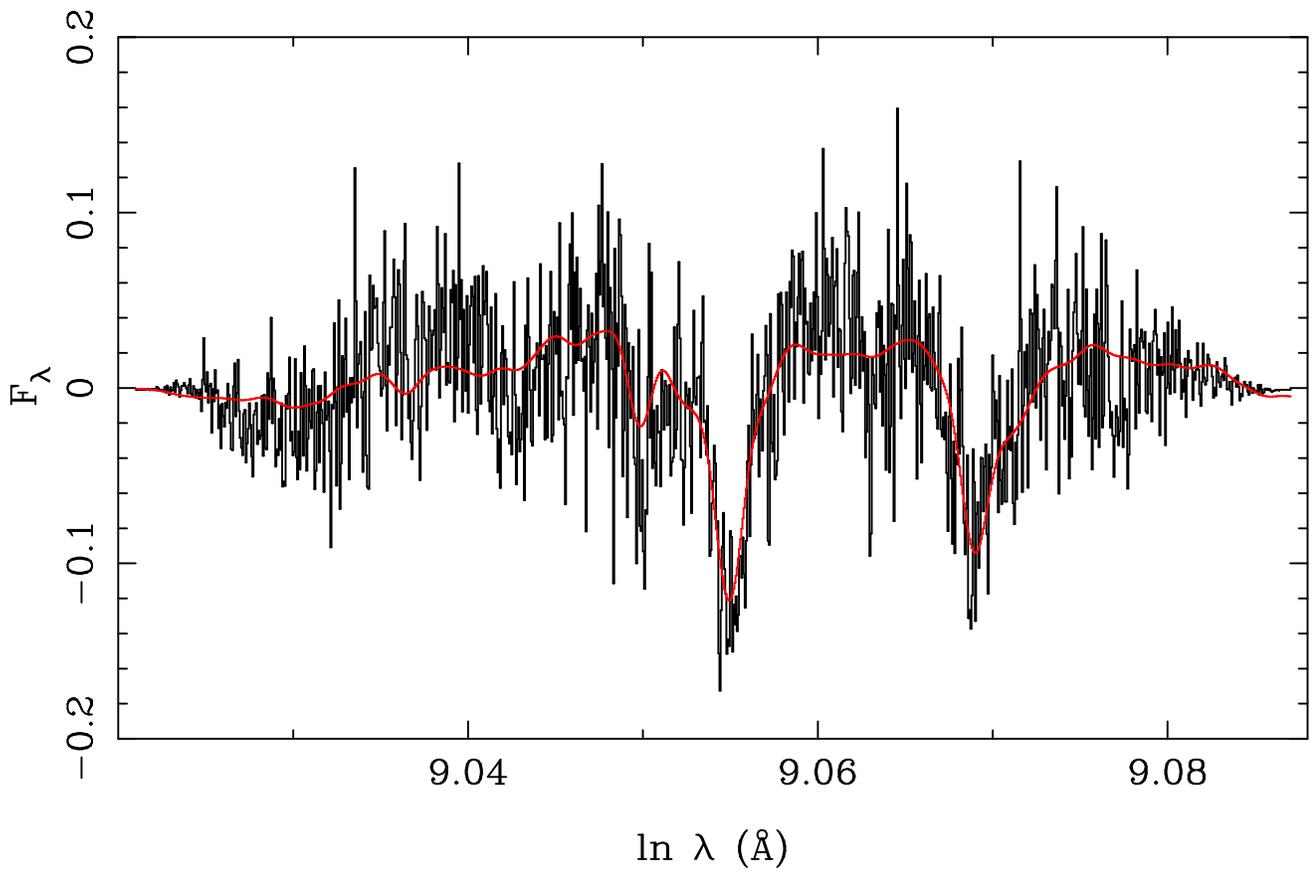
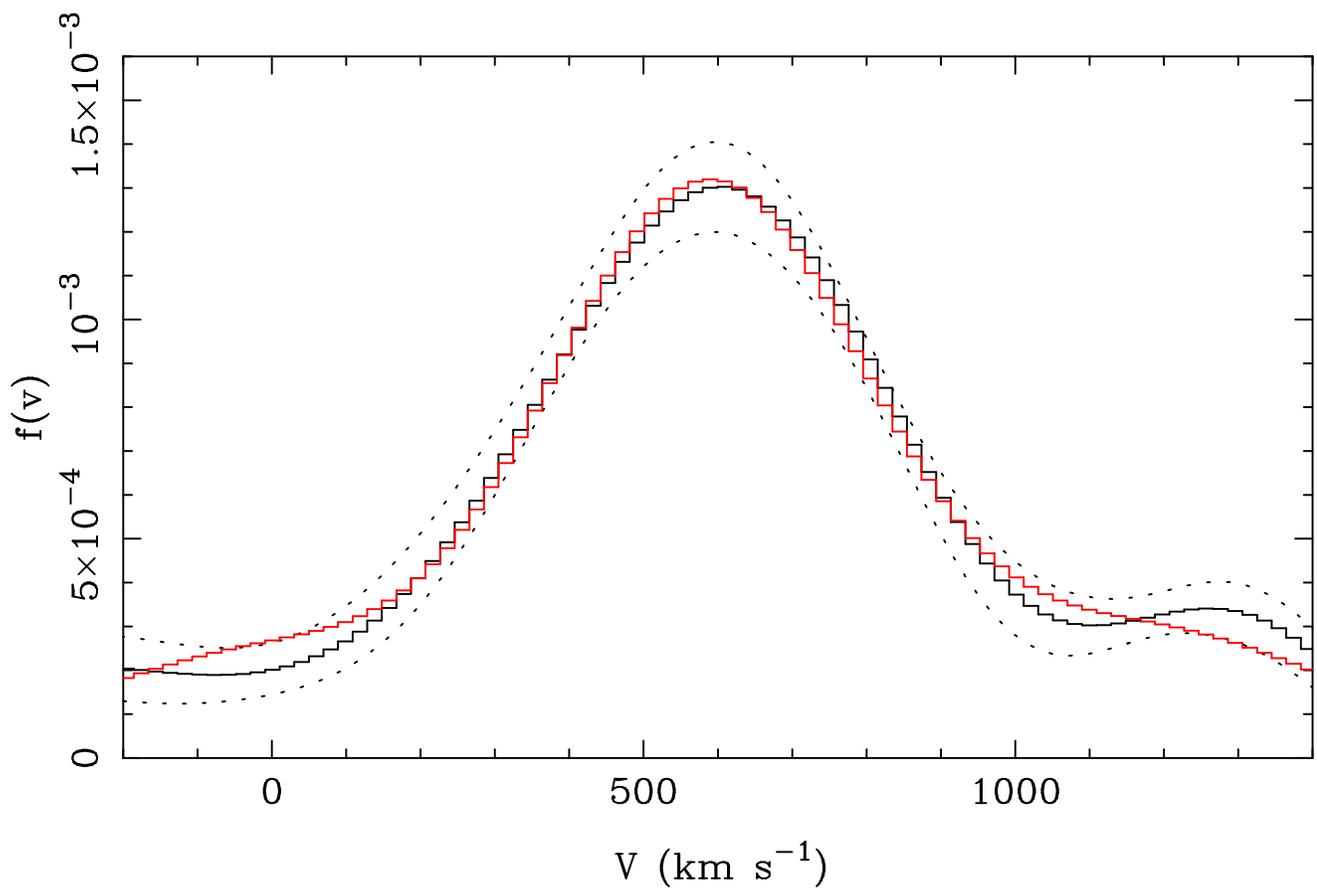
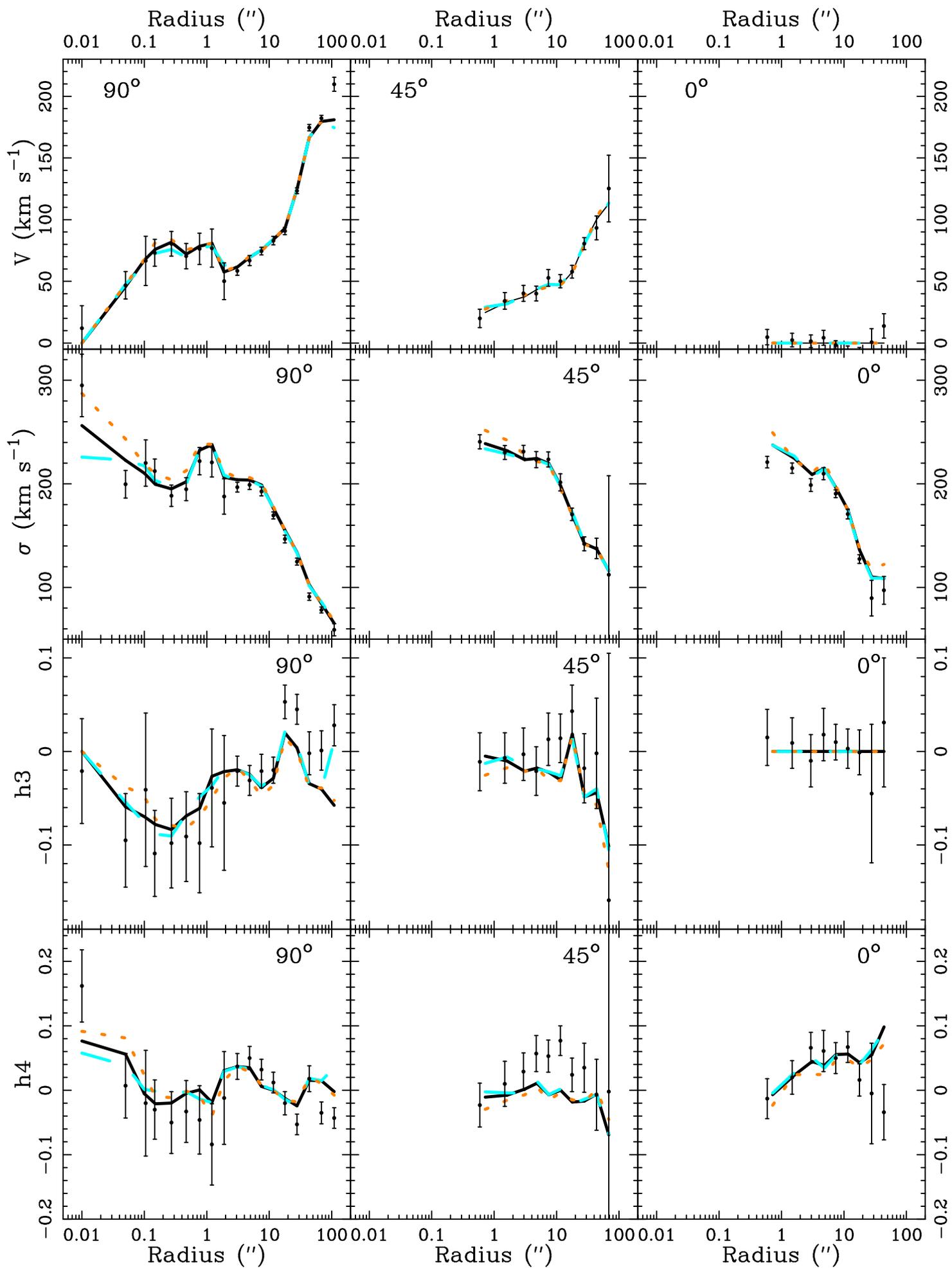
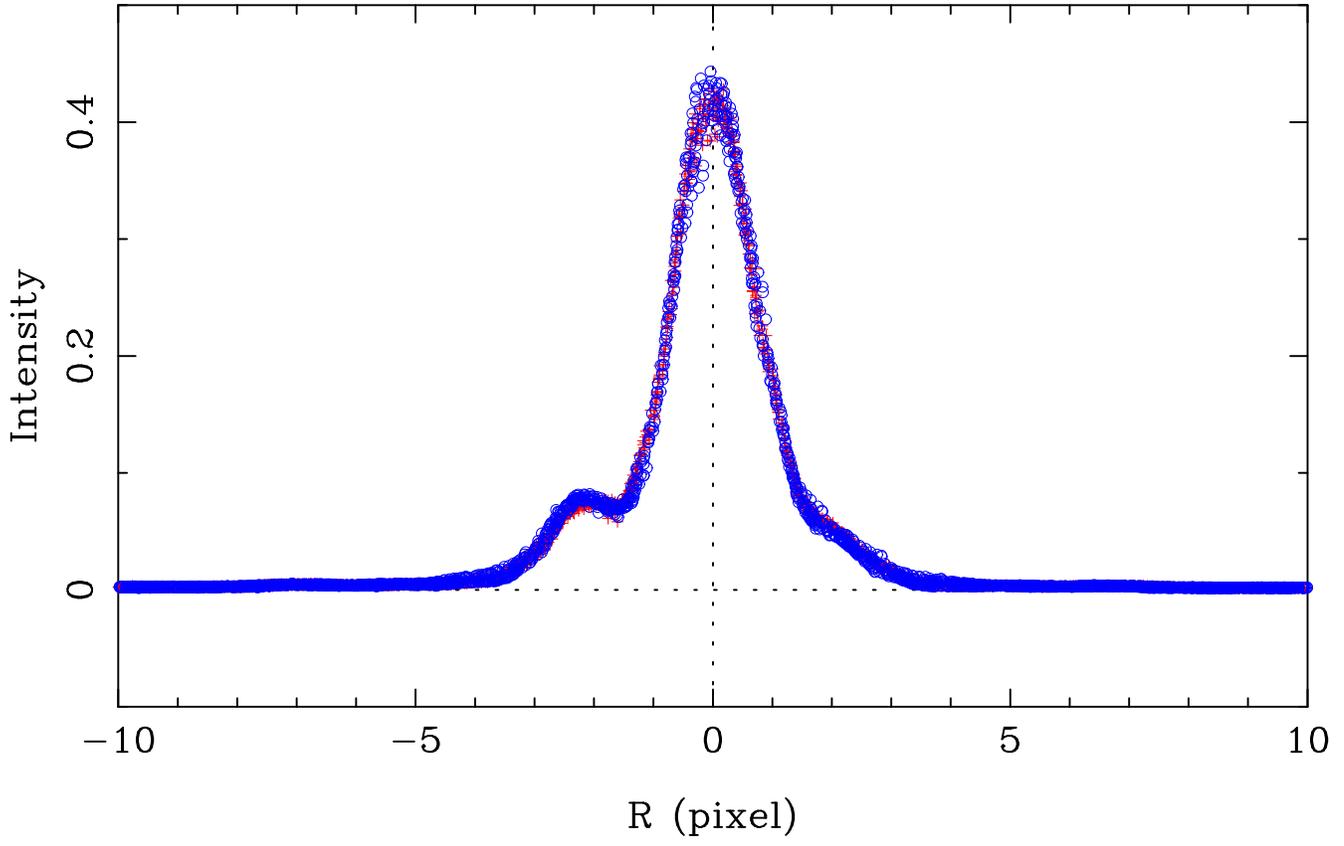
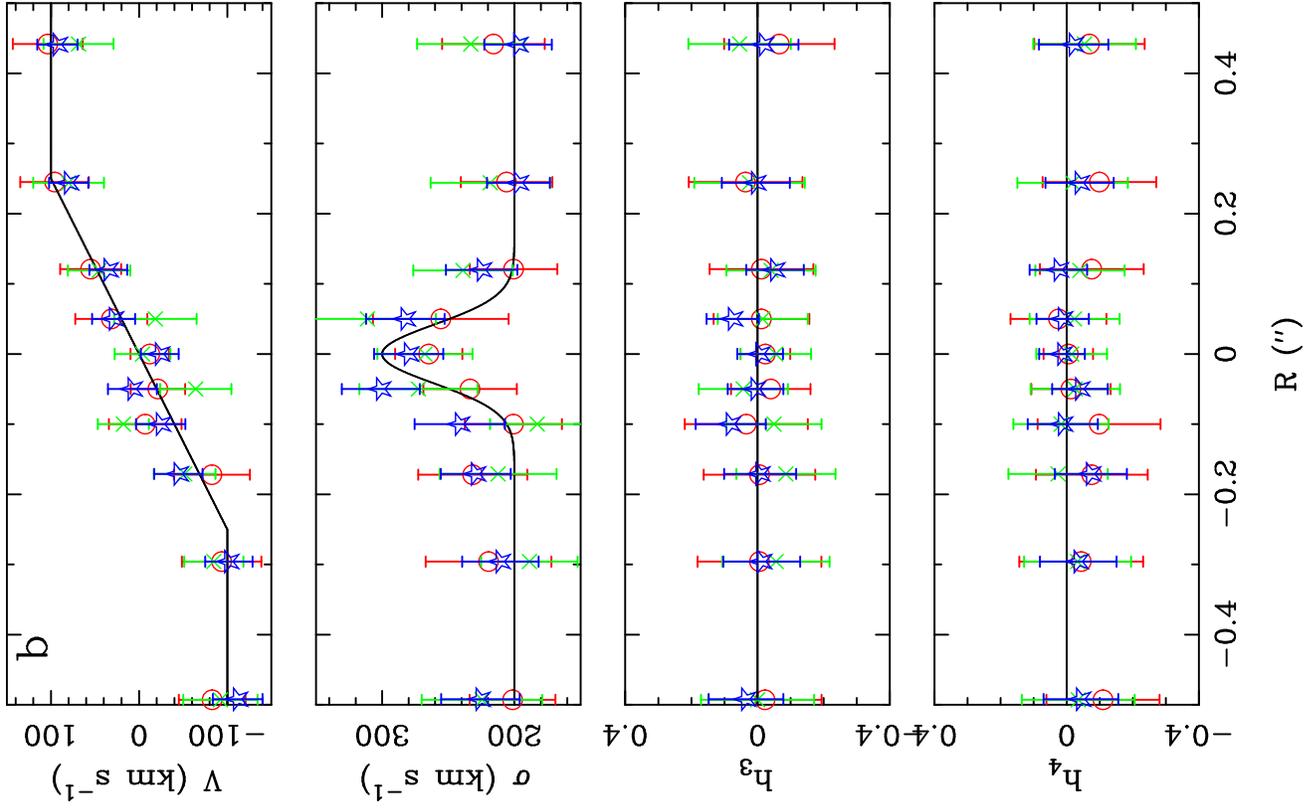
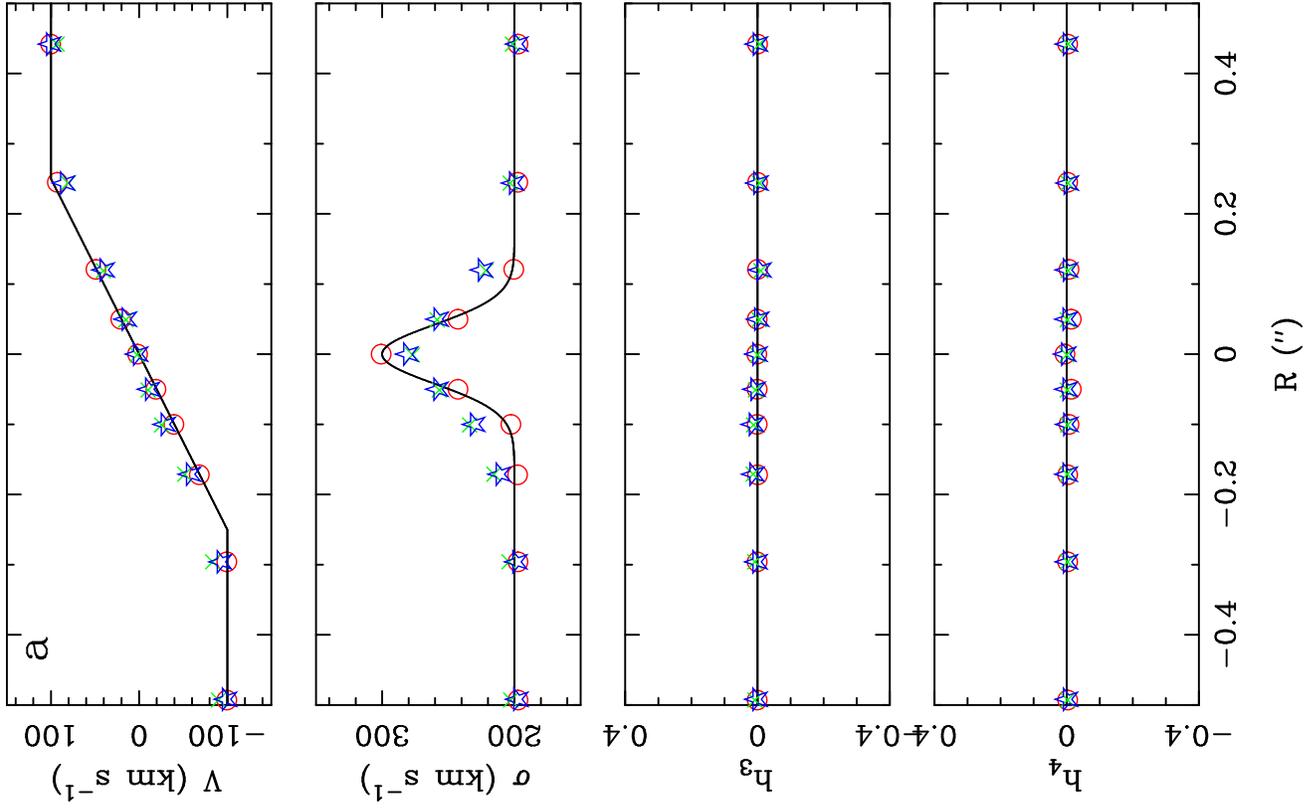
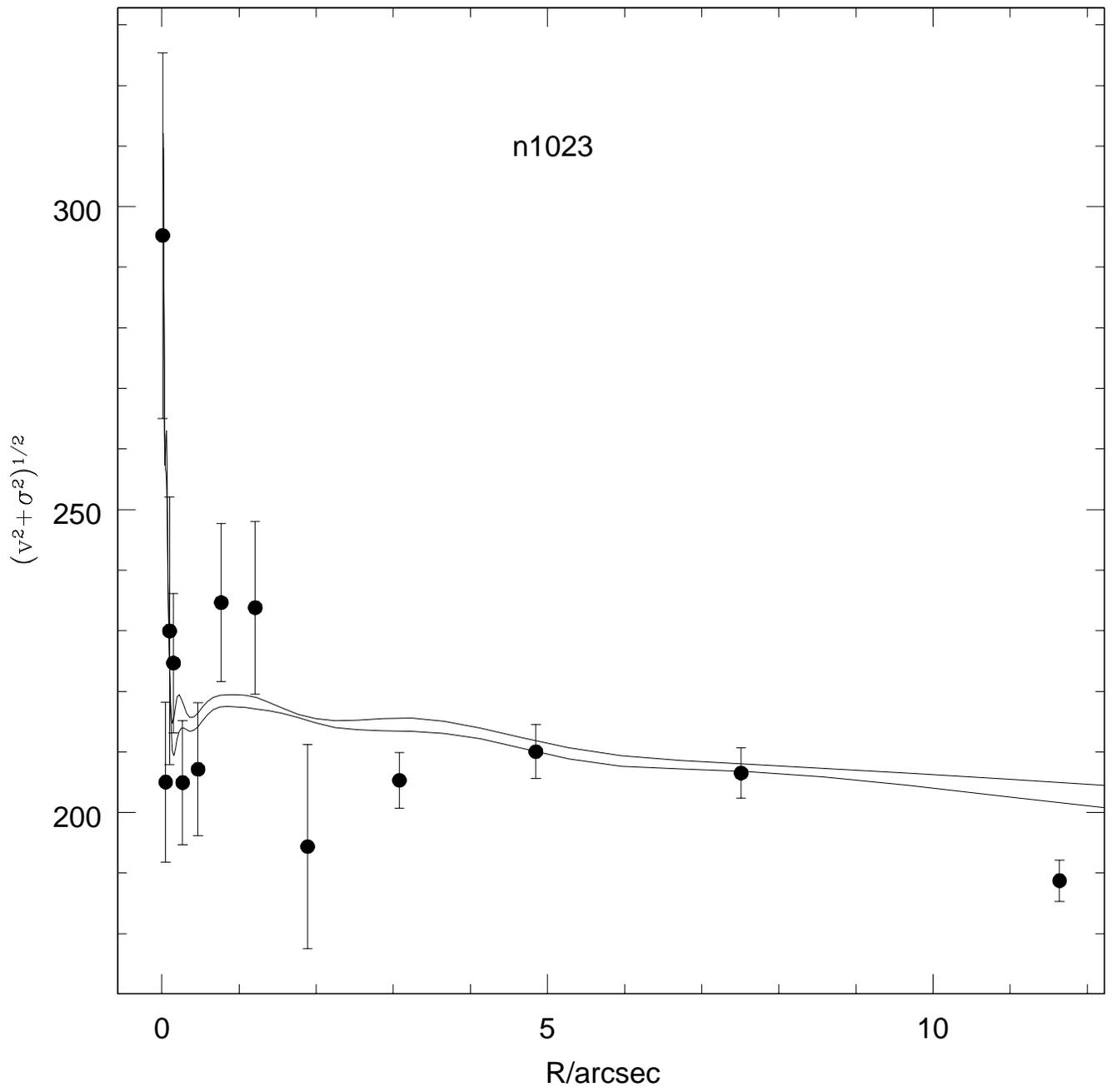
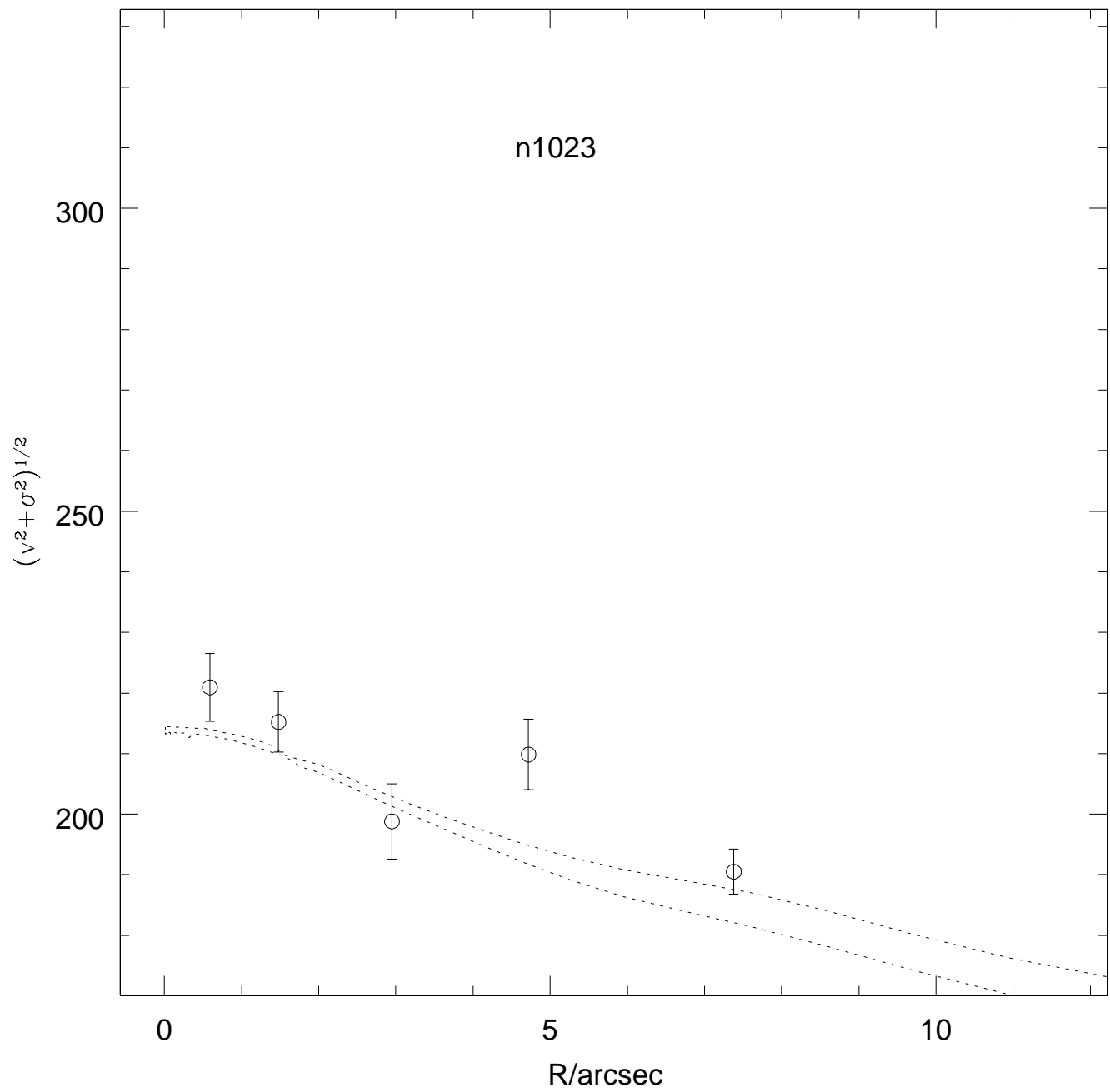
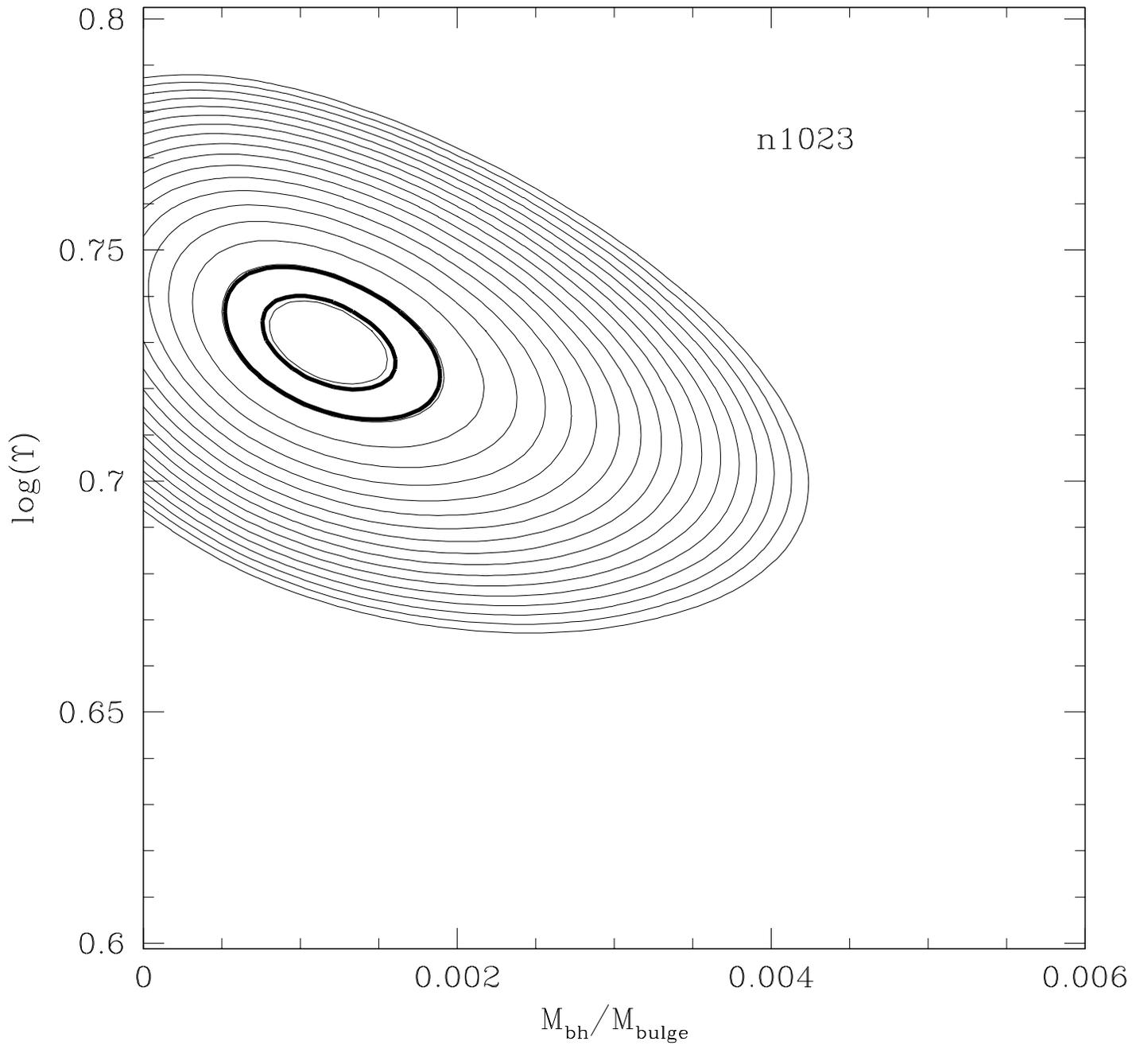
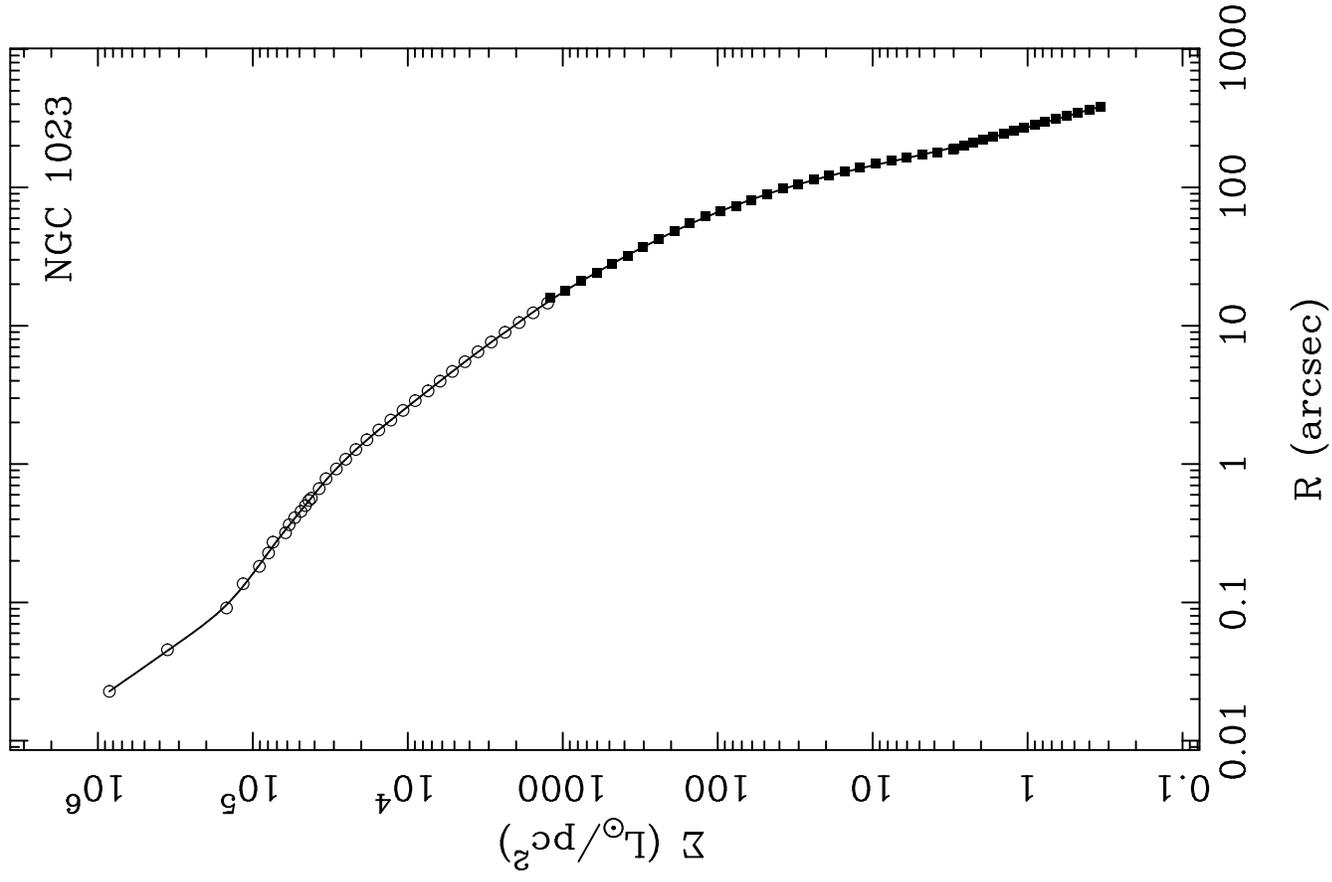
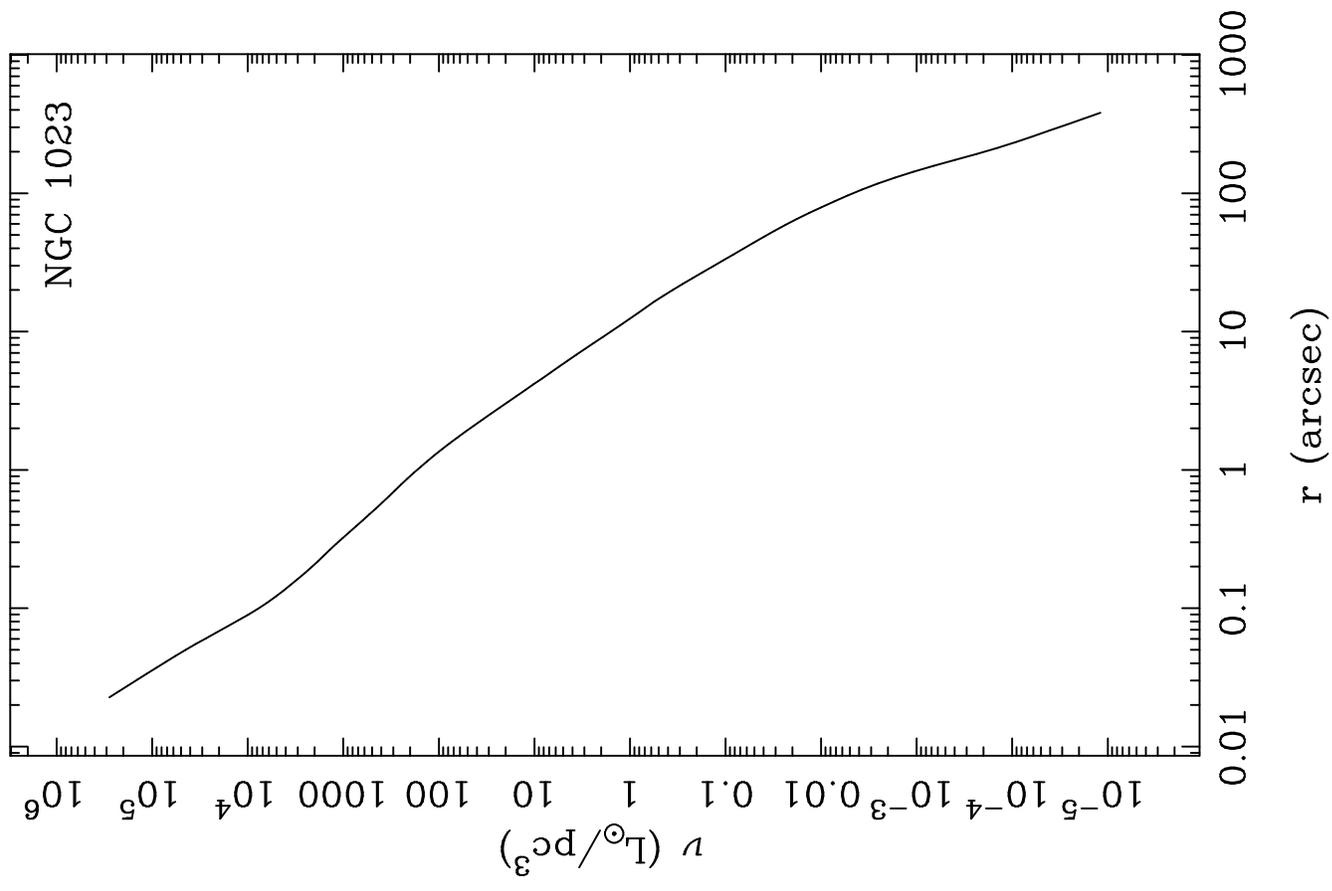
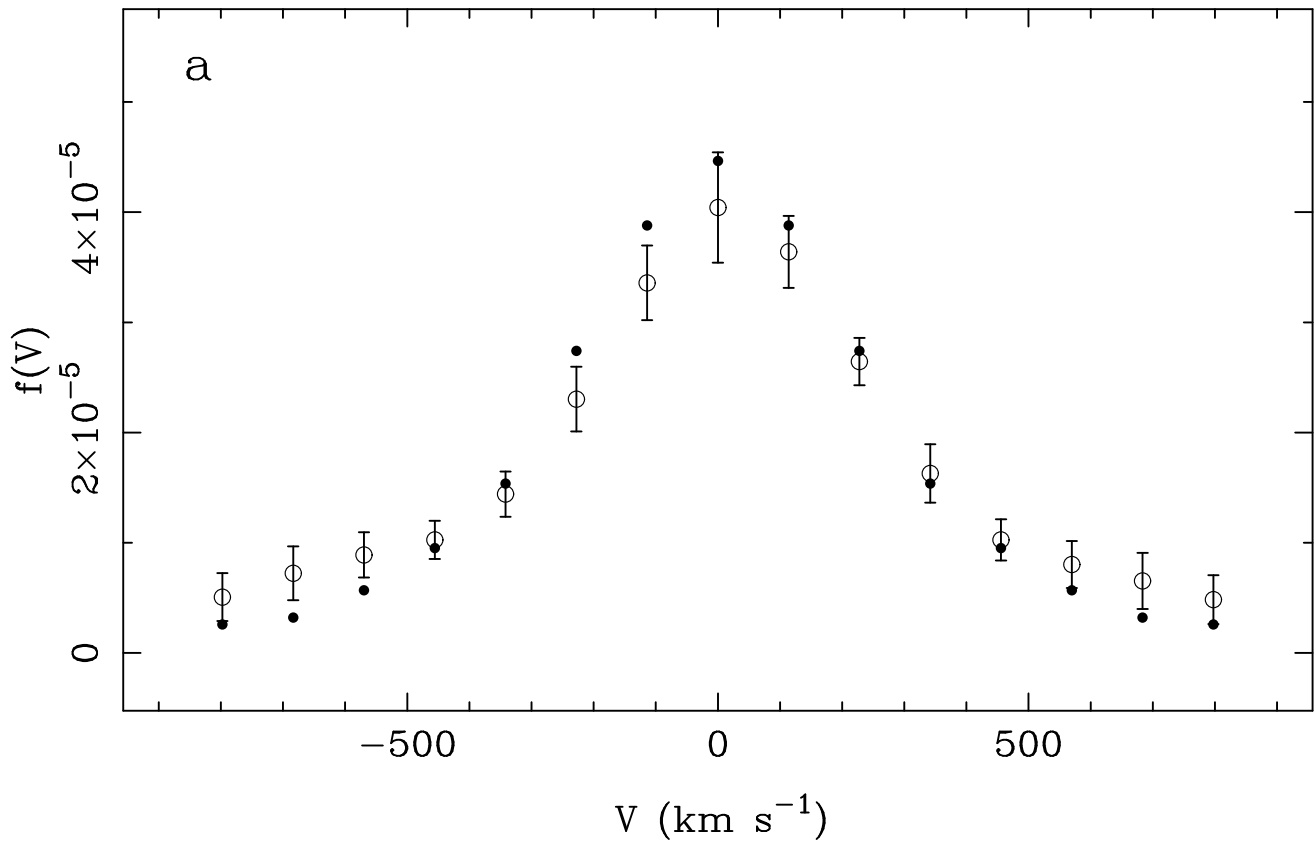
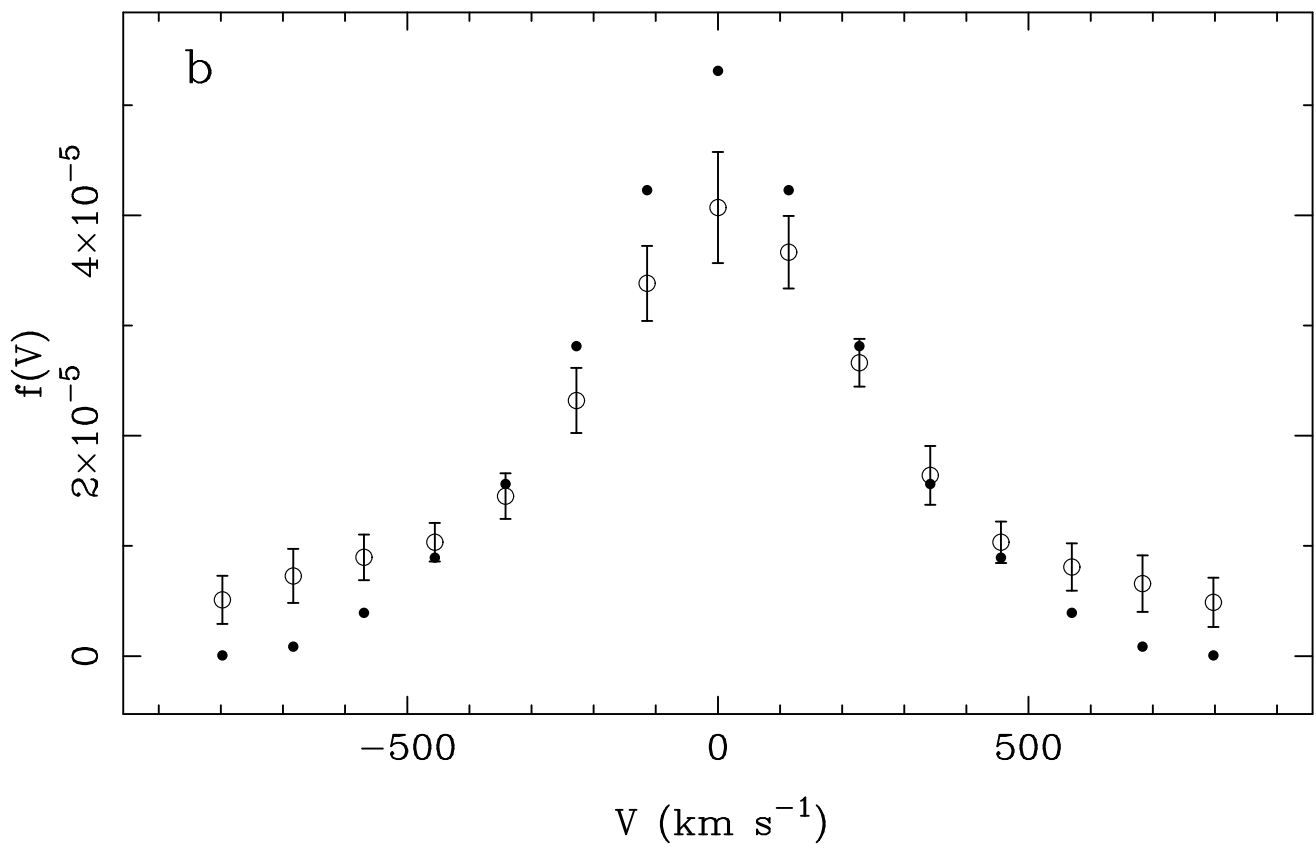
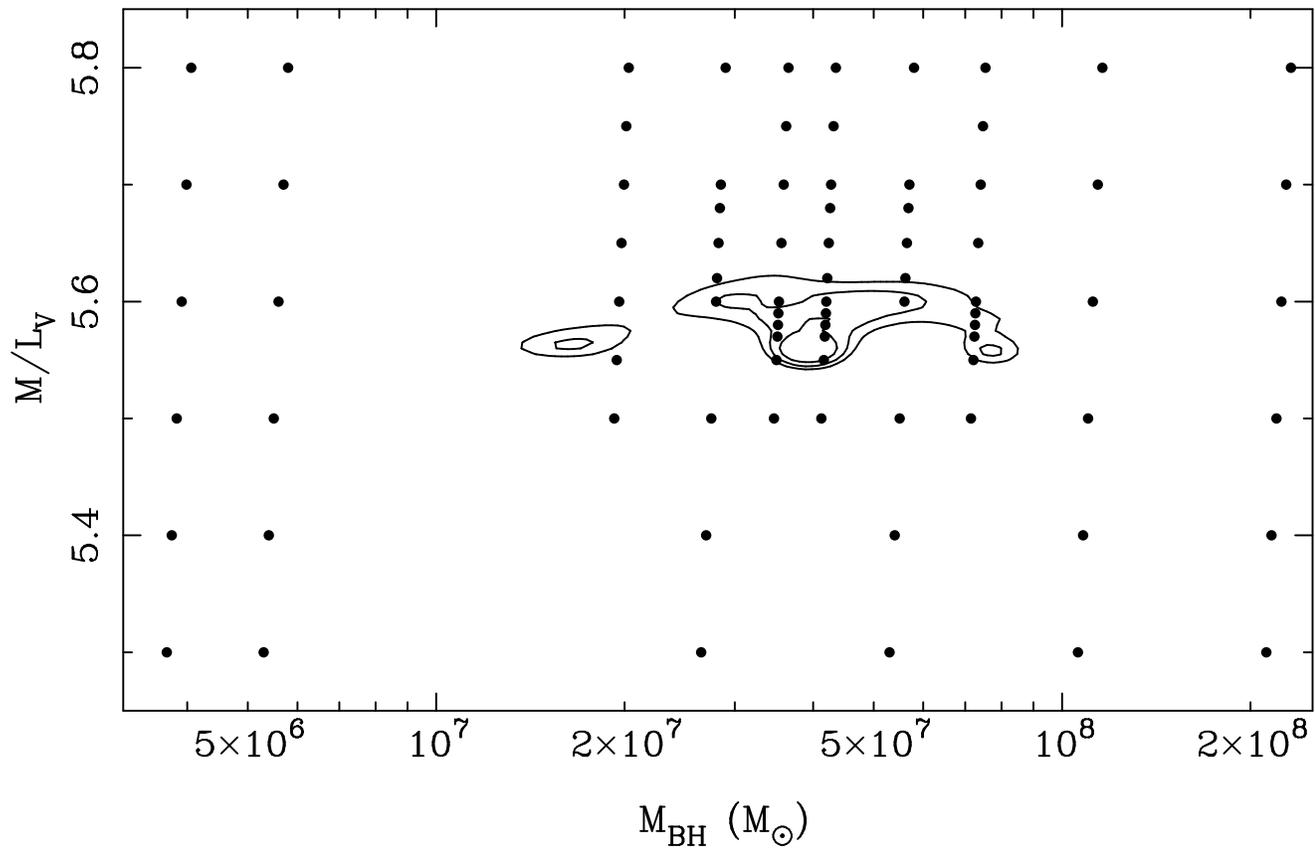
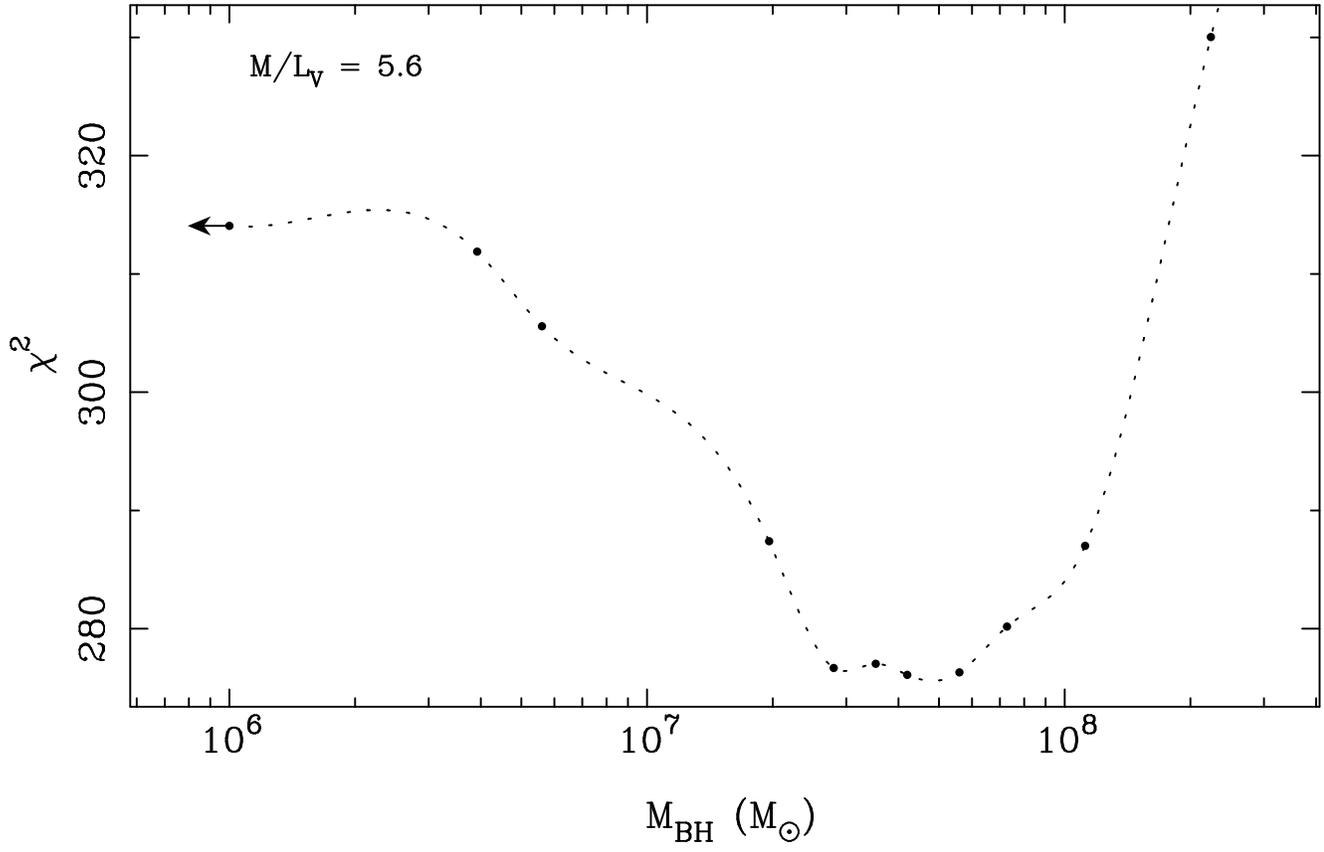
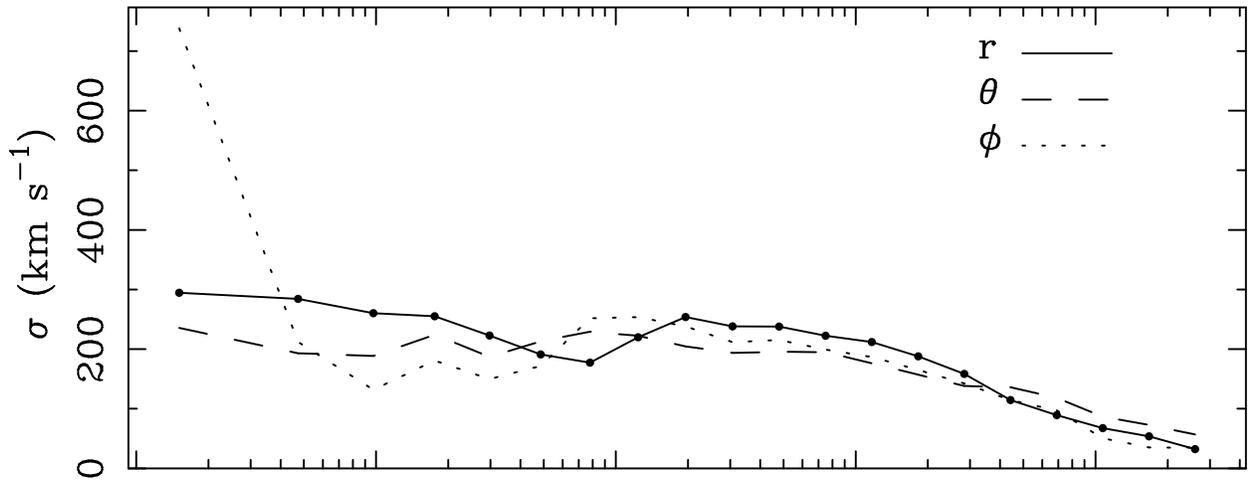
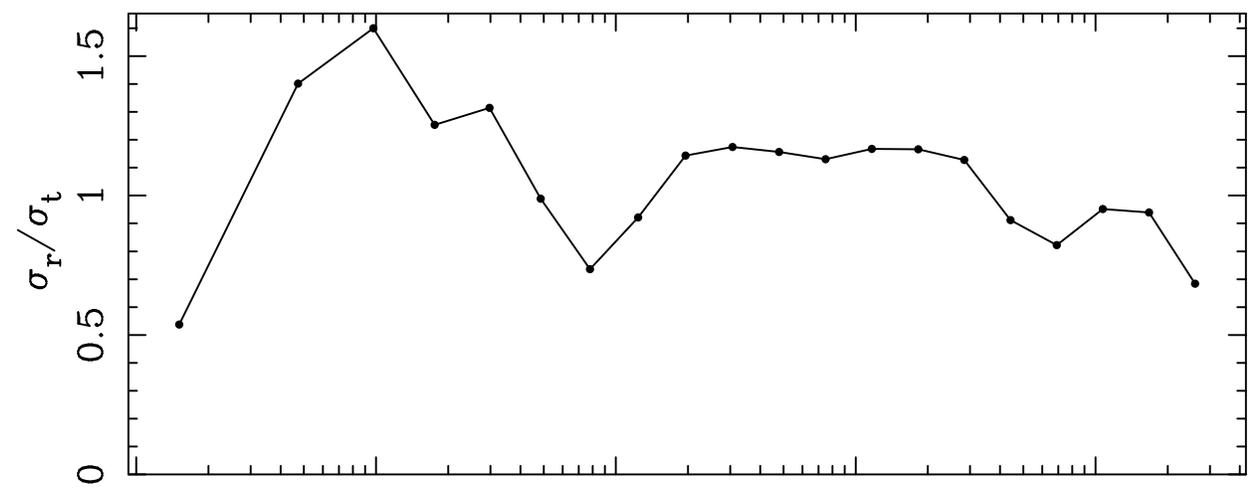
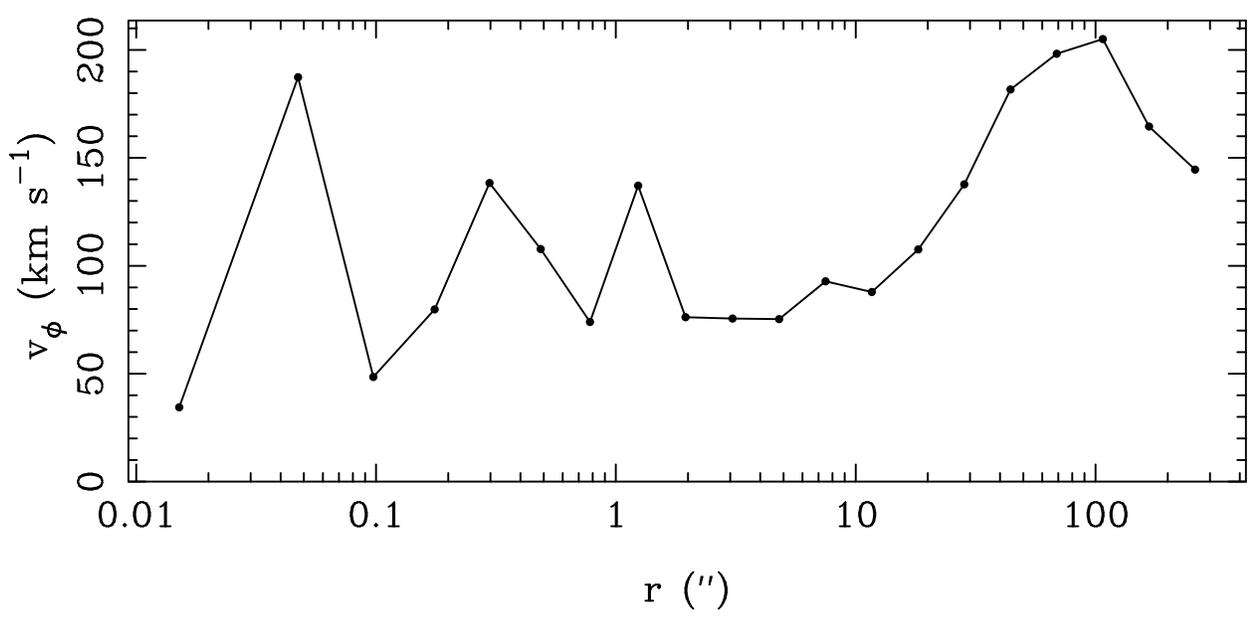
\end{document}